\documentclass[a4paper,11pt]{article}
\usepackage{jheppub}

\usepackage[most]{tcolorbox}

\usepackage{amsmath}

\usepackage{esint}
\usepackage{babel}
\usepackage{color}
\usepackage{enumitem}
\usepackage{subfigure}

\usepackage{float}
\usepackage{bbold}
\usepackage{amsfonts}
\usepackage{amssymb}
\usepackage{amsthm}
\usepackage{enumitem}
\usepackage{shuffle}
\usepackage{comment}
\usepackage{cancel}
\usepackage{blkarray}
\usepackage{multirow}
\usepackage{subfigure}
\usepackage{longtable}
\usepackage{savesym}
\savesymbol{div}
\usepackage{physics}
\restoresymbol{TXF}{div}
\usepackage{tikz}
\usepackage{tikz-3dplot}
\usepackage{xcolor}
\usetikzlibrary{patterns, decorations.markings, arrows}
\usetikzlibrary{arrows.meta}
\usepackage{mathabx}
\usepackage{mathtools}
\usepackage[export]{adjustbox}

\usetikzlibrary{snakes}
\usetikzlibrary{calc}
\usetikzlibrary{decorations}

\preprint{MPP-2025-88, USTC-ICTS/PCFT-25-17}

\title{Hexagonal Wilson loop with Lagrangian insertion at
two loops in $\mathcal{N}=4$ super Yang-Mills theory}
\author[a]{Sérgio Carrôlo,}
\author[b]{Dmitry Chicherin,}
\author[a]{Johannes Henn,}
\author[a]{Qinglin Yang,}
\author[c,d,e]{Yang Zhang}

\affiliation[a]{Max-Planck-Institut für Physik, Werner-Heisenberg-Institut, Boltzmannstr. 8, 85748 Garching, Germany}

\affiliation[b]{LAPTh, Université Savoie Mont Blanc, CNRS, B.P. 110, F-74941 Annecy-le-Vieux, France}

\affiliation[c]{Interdisciplinary Center for Theoretical Study, University of Science and Technology of China,
Hefei, Anhui 230026, China}

\affiliation[d]{Peng Huanwu Center for Fundamental Theory, Hefei, Anhui 230026, China}

\affiliation[e]{Center for High Energy Physics, Peking University, Beijing 100871, People’s Republic of China}

\emailAdd{scarrolo@mpp.mpg.de}
\emailAdd{chicherin@lapth.cnrs.fr}
\emailAdd{henn@mpp.mpg.de}
\emailAdd{qlyang@mpp.mpg.de}
\emailAdd{yzhphy@ustc.edu.cn}

\abstract{In this work, we compute the two-loop result of the null hexagonal Wilson loop with a Lagrangian insertion in planar, maximally supersymmetric Yang-Mills theory via a bootstrap approach. Normalized by the null polygonal Wilson loop itself, the integrand-level result of this observable corresponds to the logarithm of the six-point three-loop amplitude in this theory, while its integrated result is conjectured to match the maximal transcendental part of the six-point three-loop all-plus  amplitude in pure Yang-Mills theory. 
Our work builds on two recent advances. On the one hand, the set of leading singularities relevant to this observable was recently classified.
On the other hand, the relevant space of special functions that may in principle accompany these leading singularities was determined at two loops and for six particles by a dedicated Feynman integral calculation.
These two ingredients serve as the foundation of our bootstrap ansatz.
We fix all indeterminates in this ansatz by imposing physical constraints, such as symmetries, absence of spurious divergences, and correct behavior in soft and collinear limits.
Finally, we discuss and verify certain physical properties of our symbol result, including physical singularities, behavior under multi-Regge limit, as well as Steinmann relations between symbol entries. The latter relations
are motivated by the correspondence to all-plus amplitudes in pure Yang-Mills theory, and successfully checking them constitutes a consistency check of this conjectured correspondence.}

\begin{document}

\maketitle

\newpage

%%%%%%%%%%%%% INTRO %%%%%%%%%%%%%%%%

\section{Introduction}

Since
they were introduced
in~\cite{Wilson}, Wilson loops have been incredibly useful tools in high-energy theory research, and have allowed to probe and connect qualitatively different physics. 
This fact is even more remarkable when we study these objects in planar 
maximally supersymmetric Yang-Mills (sYM) theory \cite{Arkani-Hamed:2008owk} (for recent reviews, cf. \cite{Arkani-Hamed:2022rwr,Travaglini:2022uwo}), 
where light-like polygonal Wilson loops were found to be dual to maximally-helicity-violating (MHV) scattering amplitudes in the same theory \cite{Alday:2007hr,Drummond:2007aua,Brandhuber:2007yx} (see also supersymmetric versions~\cite{Mason:2010yk,Caron-Huot:2010ryg} which have interplay with %$\mathcal{N}=4$ 
sYM amplitudes of all helicity sectors). Being a relationship between conformally-invariant observables, this duality reveals many significant properties of scattering amplitudes, such as dual conformal (DCI) \cite{Drummond:2008vq} and Yangian invariance \cite{Drummond:2009fd}. 
Taking these symmetries as a starting point, tremendous progress has been made in the past decades in the study of scattering amplitudes in sYM in perturbation theory. Examples include a complete understanding of on-shell diagrams in this theory in terms of the Positive Grassmannian, also revealing cluster algebraic structures in loop integrands \cite{Arkani-Hamed:2012zlh}. 
Furthermore, 
all-loop on-shell recursion relations were discovered \cite{Arkani-Hamed:2010zjl}, as well as
 a geometric description via the Amplituhedron \cite{Arkani-Hamed:2013jha,Arkani-Hamed:2013kca}.

There is also tremendous progress for integrated quantities at loop level. For example, the functional form of four- and five-particle amplitudes is known to all loop orders thanks to dual conformal Ward identities \cite{Drummond:2007au}, in agreement with the Bern-Dixon-Smirnov conjecture \cite{Bern:2005iz} and with the AdS/CFT prescription for computing scattering amplitudes at strong coupling \cite{Alday:2007hr}.
The dual conformal Ward identities allow for non-trivial functions of conformal cross-ratios to appear starting from six particles.
These are best described via infrared-finite ratio and remainder functions \cite{Drummond:2008vq}.
In this direction, 
many advanced methods have been developed to study these functions, including: the \textit{symbol} for multi-polylogarithmic functions \cite{Goncharov:2010jf,Duhr:2011zq}; the cluster algebras relevant to the function spaces \cite{Golden:2013xva}; the Steinmann /cluster algebra bootstrap strategy \cite{Dixon:2011pw, Caron-Huot:2016owq,Dixon:2016nkn,Drummond:2018caf,Caron-Huot:2019vjl,Caron-Huot:2019bsq,Caron-Huot:2020bkp}; the $\bar{Q}$-anomaly equations \cite{Caron-Huot:2011zgw,Caron-Huot:2011dec}, along with many others.
Moreover, leveraging the Wilson loop picture, operator product expansion techniques have been developed \cite{Alday:2010ku,Basso:2013vsa,Basso:2013aha} that allow for a non-perturbative description of these quantities in the near-collinear regime.

Despite all this progress, few of the above results have explicitly connected the wealth of integrand-level results to the loop functions, as the methods used for obtaining the latter are mostly indirect, or rely on certain conjectures.
Going from loop integrands to integrated results directly, \textit{i.e.}, evaluating the Feynman loop integrals, still remains a major bottleneck. 
Any progress in this direction would not only help us to obtain more data points for integrated results of scattering amplitudes, but it would also have significant applications in more general physical theories that are experimentally accessible, such as the computation of scattering amplitudes in Quantum Chromodynamics (QCD). 

Readers interested in QCD applications might object that specialized techniques for dual conformally invariant scattering amplitudes are of little use to theories that do not possess this symmetry. There are two answers to this concern. First, one may always use dual conformal symmetry to fix a certain frame, and in this way relate dual conformal Feynman integrals to non-dual-conformal ones, as \textit{e.g.} in \cite{Broadhurst:1993ib}.
Second, there is a specific object in sYM theory, which we describe presently, which was observed to be closely related to pure Yang-Mills amplitudes, and in fact appears to depend on the same function space. It therefore appears to be a perfect playground for developing novel methods for loop integration, with potential relevance to QCD.
We consider the following observable,
\begin{align}
    F_n(x_1,\ldots,x_n;x_0) := \pi^2 \frac{\vev{W_n[x_1,\ldots,x_n] {\cal L}(x_0)}}{\vev{W_n[x_1,\ldots,x_n]}} \,,
\end{align}
which we refer to as \textit{Wilson loop with a Lagrangian insertion}. It has been studied in a series of previous works~\cite{Alday:2011ga,Alday:2012hy,Alday:2013ip,Engelund:2011fg,Engelund:2012re,Henn:2019swt,Chicherin:2022bov,Chicherin:2022zxo,Bargheer:2024hfx,Chicherin:2024hes}. 
In the equation above, $\vev{W_n}$ is the vacuum expectation value of a null polygonal Wilson loop in the fundamental representation of the color group $SU(N_c)$, and the operator ${\cal L}$ is the
{\em chiral on-shell Lagrangian}~\cite{Eden:2011yp} of ${\cal N} =4$ sYM,
which is a protected operator of conformal dimension four. The expansion of this object in perturbation theory enjoys many ideal properties. First, as a consequence of the duality between Wilson loops and scattering amplitudes, its integrand is given by the logarithm of the $n$-point scattering amplitude, treating $x_0$ as an extra loop integration. Furthermore, infrared divergences are exactly canceled when taking the ratio, indicating that for each loop-order we only need to deal with finite integration.

Surprisingly, Wilson loops with a Lagrangian insertion in sYM have been found to be related to
all-plus amplitudes in non-supersymmetric Yang-Mills theory.
More precisely, relationship has been observed between the leading transcendental part of the all-plus pure Yang-Mills amplitudes 
and the expressions in sYM theory \cite{Chicherin:2022bov,Chicherin:2022zxo}. 
Moreover, if one uses the dual conformal symmetry to go to the kinematical frame $x_0\to\infty$, then the kinematical dependence corresponds to that of a general $n$-point scattering process.
These observations suggest that understanding the Wilson loop observable better may also advance understanding of QCD amplitudes.

One way of characterizing the observable in a perturbative expansion is by employing bootstrap methods, and in particular, the {\it symbol bootstrap}. 
The idea is to make an ansatz for the observable, at some order in the perturbative expansion.
The input for such an ansatz is the knowledge of the functions that can appear (through the symbol), as well as the possible leading singularities \cite{Cachazo:2008vp,Bullimore:2009cb,Arkani-Hamed:2010pyv} accompanying these functions. 
Specifically, one writes, for the $L$-loop contribution in perturbation theory (defined as the coefficient of $(g^2)^L$, where $g^2 = g_{\rm YM}^2 N_c/(16 \pi^2)$, in the `t Hooft limit),
\begin{align}
\label{eq:symbolbootstrapintroduction}
F^{(L)} = \sum_{ij} c_{ij} R_{i} f_{j}^{(L)} \,.
\end{align}
Here $R_i$ is a set of leading singularities, $f_j$ the set of relevant symbols, and the $c_{ij} \in {\mathbb{Q}}$ are indeterminates.
Then, taking this ansatz and imposing suitable physical conditions such as symmetry constraints and appropriate behavior in kinematical limits, one can attempt to fix the final expression for the observable. 
Such a procedure showed overwhelming power in calculating six- and seven-point scattering amplitudes, as well as three- and four-point half-BPS operator form factors in $\mathcal{N}=4$ sYM theory~\cite{Dixon:2011pw,Dixon:2014xca,Dixon:2014iba,Drummond:2014ffa,Dixon:2015iva,Caron-Huot:2016owq,Dixon:2016nkn,Drummond:2018caf,Caron-Huot:2019vjl,Caron-Huot:2019bsq,Caron-Huot:2020bkp,Dixon:2020bbt,Dixon:2022rse, Dixon:2022xqh,Dixon:2023kop,Basso:2024hlx}.

The symbol bootstrap relies crucially on the knowledge of both the relevant symbols $f_j$ and leading singularities $R_i$. For the cases mentioned above, the relevant symbol alphabets were conjecturally known, thanks to connections to cluster algebras.
In the case of Wilson loops with Lagrangian insertion, and likewise for corresponding scattering amplitudes in pure Yang-Mills theory, the relevant function spaces are expected to be more complicated.
Let us give some examples of the state-of the art in QCD amplitudes. There is a host of interesting results for four- and five-particle scattering amplitudes, cf. references \cite{Caola:2020dfu,Caola:2021izf,Caola:2021rqz,Abreu:2018zmy,Abreu:2019odu,Abreu:2021oya,Badger:2013gxa,Badger:2015lda,Badger:2018enw,Badger:2019djh,Badger:2023eqz,Agarwal:2023suw,DeLaurentis:2023nss,DeLaurentis:2023izi,Henn:2016jdu,Abreu:2018aqd,Chicherin:2018yne}. Very recently, computations of the six-point two-loop planar function space~\cite{Henn:2025xrc,Abreu:2024fei} via the canonical differential equations method \cite{Henn:2013pwa}, have opened up whole new possibilities, 
which inspire us to consider physical observables that localize in this new function space. 

Moreover, a recent paper \cite{Brown:2025plq} proved an important conjecture regarding the leading singularities appearing in Wilson loops with a Lagrangian insertion at all loop orders. 
Hence, equipped with these tools, we are in a perfect position to apply bootstrap methods to a hexagonal Wilson loop with Lagrangian insertion at two-loop order. 
This is what we do in the present paper.
We note that this will be 
the first computed observable living in the newly derived planar hexagon two-loop function space, and as such may give a first glimpse of what to expect for QCD amplitudes at this order in perturbation theory.

It may be interesting to compare the premise of the bootstrap ansatz, eq. \eqref{eq:symbolbootstrapintroduction} for six-point ratio/remainder functions to the case of Wilson loops with Lagrangian insertion. In the former case, the alphabet consists of nine letters, depending on three independent variables, and there are either one or five 
independent leading singularities, depending on the case considered. In the latter case, there are $244$ dimensionless alphabet letters \cite{Henn:2025xrc,Abreu:2024fei} that depend on seven independent variables, and there are 20 leading singularities.
This counting gives an idea of the typical increase in complexity when going from sYM amplitudes to pure Yang-Mills amplitudes.

There is a simplification that we can profit from thanks to the explicit Feynman integral results obtained in refs. \cite{Henn:2025xrc,Abreu:2024fei}, compared to the usual symbol bootstrap procedure. Usually one starts from the symbol alphabet, and constructs integrable symbols of a certain length (corresponding to the expected transcendental weight). For $245$ alphabet letters, this step in itself could become challenging. However, we can bypass this and read off directly from the results of refs. \cite{Henn:2025xrc,Abreu:2024fei} the relevant weight-four functions. A similar approach was taken in ref. \cite{Guo:2021bym} in the context of bootstrapping certain form factors.

The paper is organized as follows. In section~\ref{sec: review}, we give a brief review of the ingredients needed for the bootstrap, including six-point kinematics, leading singularity structures
of the observable, basics of symbol bootstrap strategy and six-point two-loop planar function space. 
In section \ref{sec: bootstrap and discussion}, we show how a suitable behavior in specific physical limits allows for the observable to be completely fixed.  Then we discuss the bootstrapped result, including organization of the symbol result and singularity information.  We conclude this section by showing that the duality to all-plus amplitudes guarantees that the respective all-plus amplitude automatically satisfies the Steinmann relations. Discussion and outlook are offered finally in section \ref{sec: outlook}. In Appendix~\ref{app:fourfive}, we record some previously known results at four and five points, which serve as boundary conditions for our calculation. The multi-Regge limit of the result is discussed in Appendix~\ref{app:remarks}.

\section{Preliminaries}
\label{sec: review}

\subsection{Six-particle kinematics}

In this work, we consider an observable whose kinematical dependence is the same as that of a scattering process for six massless particles. Consider six outgoing four-momenta subjected to the conditions of both total momentum conservation and on-shell conditions, which read
\begin{equation}
    \sum_{i=1}^6p_i=0\,, \quad p_i^2=0 \, , \quad \forall\ {i\in \{1,\dots,6\} } \, .
\end{equation}
In general dimensions, there are nine linearly-independent Mandelstam variables. We choose the basis comprised of the following planar variables,
\begin{equation}\label{eq:sdef6}
    \{s_{i,i{+}1}=(p_i+p_{i+1})^2 \}_{\ i=1,\cdots,6}\, \cup\ \{ s_{i,i+1,i+2}=(p_i{+}p_{i{+}1}{+}p_{i{+}2})^2 \}_{i=1,2,3} \, .
\end{equation}
Following the notation above, we also conveniently define $s_A:=(\sum_{i\in A}p_i)^2$ for the rest of the work. In the present paper, we work in four dimensions. In this case, the six-particle kinematics is further constrained by the Gram-determinant condition,
\begin{equation}
    \mathcal{G}(p_1,p_2,p_3,p_4,p_5) \equiv \det G(p_1,p_2,p_3,p_4,p_5) = 0\,,
\end{equation}
where the Gram matrix $G(v_1,\ldots v_k)$'s $(i,j)$-th entry is $2v_i \cdot v_j$. Thus, as a consequence, the number of independent Mandelstam variables is reduced to eight.

In addition to the parity-even invariants defined above, one can also construct parity-odd Lorentz-invariants using the antisymmetric Levi-Civita tensor as
\begin{equation}
    \label{eq:epsijkl1}         
    \epsilon(i,j,k,l):=4i\varepsilon_{\mu\nu\rho\sigma}p_i^\mu p_j^\nu p_k^\rho p_l^\sigma \, , \, \forall\ {i,j,k,l \in \{1,\dots,6\}} \, .
\end{equation}
Moreover, notice that the product of two $\epsilon(i,j,k,l)$ is given by a parity-even generalized Gram determinant, which for the case of $\epsilon(i,j,k,l)^2$ simplifies to
\begin{equation}\label{eq:epsijkl}
    \epsilon(i,j,k,l)^2=\mathcal{G}(p_i,p_j,p_k,p_l) \, .
\end{equation}

\paragraph{Spinor-helicity variables, dual coordinates and momentum twistors.}

The use of Mandelstam variables introduces square-root like expressions, such as $\epsilon(i,j,k,l)$. Furthermore, the momentum conservation and on-shell conditions impose, as we have seen, extra constraints on these variables that need to be taken into account when simplifying expressions. Instead of this, it is useful to work with variables that take care of these constraints automatically. 

Starting with the on-shell condition, we can trivialize it by using the map
\begin{equation}
(p_i)_\mu \rightarrow p_{\alpha \dot{\beta}} \equiv (p_i)_\mu (\sigma^{\mu})_{\alpha\dot{\beta}}=(\lambda_i)_{\alpha}(\tilde{\lambda}_i)_{\dot{\beta}} \,
\label{eq:spinHel}
\end{equation}
where  $\sigma^\mu$ is a four vector consisting of the identity and the Pauli matrices, $\sigma^\mu = (\mathbb{1},\sigma^i)$. The RHS is obtained as a consequence of the on-shell condition for $p_i$, which forces the determinant of $p_{\alpha \dot{\beta}}$ to vanish. The variables $\lambda_i$ and $\tilde{\lambda}_i$ are called  the \textit{spinor-helicity} variables, whose little-group-covariant inner products are
\begin{equation}\label{eq:twobrackets}
    \langle ab\rangle=\varepsilon_{\alpha\beta}\lambda_a^\alpha\lambda_b^\beta \, , \quad  [ab]=\varepsilon_{\dot{\alpha}\dot{\beta}}\tilde{\lambda}_a^{\dot{\alpha}}\tilde{\lambda}_b^{\dot{\beta}} \, .
\end{equation}
In addition to this, in order to trivialize the momentum-conservation condition, we begin by introducing {\it dual coordinates} $x_i\in \mathbb{R}^4$ defined as
\begin{equation}
    p_i=x_{i+1}-x_i\,,\qquad x_7 \equiv x_1 \, ,
\end{equation}
where momentum conservation has been substituted by the last equality, which ensures that the six $x_{i}$'s form a closed polygon. In these variables, the Mandelstam invariants can be written as
\begin{equation}
    \label{eq:sdef62}
    s_{i,i{+}1}= x_{i,i+2}^2 \, , \quad s_{i,i+1,i+2}=x_{i,i+3}^2
\end{equation}
where $x_{i,j}^2:=(x_i - x_j)^2$. Now, putting together the on-shell condition with momentum conservation, we can finally define \textit{momentum twistor} variables $Z_i^I\in\mathbb{P}^3$ as 
\begin{equation}
    {Z}_i^I:=(\lambda_i^{\alpha},x_{i}^{\dot{\beta}\gamma}{\lambda}_{i,\gamma})^I \, , \quad I= \{\alpha = 1,2, \dot{\beta}=1,2 \} = 1,\cdots,4 \, .
\end{equation}
These were originally defined in~\cite{Hodges:2009hk}, and they are of special interest as, besides trivializing momentum conservation and on-shell conditions, they also linearize the full conformal group in dual space to a simple $\text{SL}(4)$-invariance. This, in turn, is of interest for our discussion because $\mathcal{N}=4$ sYM theory is a DCI theory, \textit{i.e.}, it is conformally invariant in the dual space $x_i$. As a consequence, the conformally-invariant quantities are given by $4\times4$ determinants as
\begin{equation}
    \label{eq:four_bracket}
    \langle i j k l \rangle := \det(Z_i Z_j Z_k Z_l ) = \varepsilon_{IJKL} Z^{I}_i Z^{J}_j Z^{K}_k Z^{L}_l \ ,
\end{equation}
where once again $\varepsilon_{IJKL}$ is the Levi-Civita tensor in four dimensions. One can express the Mandelstam variables of equation \eqref{eq:sdef6} as
\begin{equation}
\label{eq:sdef63}
    s_{i,i{+}1}=\frac{\langle i{-}1\ i\ i{+}1\ i{+}2\rangle}{\langle i{-}1\,i\rangle\langle i{+}1\,i{+}2\rangle} \, , \quad  s_{i,i{+}1,i{+}2}=\frac{\langle i{-}1\,i\,i{+}2\,i{+}3\rangle}{\langle i{-}1\,i\rangle\langle i{+}2\,i{+}3\rangle} \, ,
\end{equation}
where we have four-brackets from \eqref{eq:four_bracket} and two-brackets from \eqref{eq:twobrackets}. The latter can be related to the former by introducing the so-called \textit{infinity bitwistor}
\begin{equation}
    I_{\infty}:=\left(\begin{matrix}
        0&0\\0&0\\1&0\\0&1
    \end{matrix}\right) \,,
\end{equation}
which represents the point at infinity in dual space. Using this, every spinor-helicity bracket can be written in terms of four-brackets as
\begin{equation}
    \langle ij\rangle = \langle ij\ I_\infty\rangle \, .
\end{equation}
Ultimately, with this map, all Mandelstam variables can then be written as ratios of momentum twistor four-brackets. 

If one then wants to compute amplitudes at a general $L$-loop order, we need to supplement the above dual variables $x_i$ with extra variables pertaining to the loops, which we call $y_\ell$, where the index $\ell$ runs from $1$ to $L$. Moreover, it can be seen that each $y_\ell$ corresponds to a line $(AB)_\ell$ in momentum twistor space.

\paragraph{Correspondence between DCI kinematics and six-point kinematics.}

Finally, it is worth mentioning that, due to the dual conformal invariance, $F_6$ depends on conformal cross-ratios built from an arbitrary $x_0$ and cusps $x_1,\ldots,x_6$ of the light-like hexagonal contour, whose multiplicatively-independent basis can be chosen as
\begin{equation}
\label{eq:8ratios}
    \left\{\frac{x_{i,i{+}2}^2x_{1,0}^2x_{3,0}^2}{x_{1,3}^2x_{i,0}^2x_{i{+}2,0}^2}\right\}_{i=2,\cdots 6}\cup\left\{\frac{x_{i,i{+}3}^2x_{1,0}^2x_{3,0}^2}{x_{1,3}^2x_{i,0}^2x_{i{+}3,0}^2}\right\}_{i=2,\cdots 4} \, .
\end{equation}
Similarly to the discussion of the six-point kinematics, in four dimensions, the distances between dual points satisfy an extra seven-point Gram determinant condition given by
\begin{equation}
   \mathcal{G}(x_0,x_1,\cdots,x_6):=0,\
\end{equation}
where $\mathcal{G}(x_0,x_1,\cdots,x_6)$ is the determinant of the symmetric matrix whose $(i,j)$-th entry is $x_{i,j}^2$, $i,j\in \{0,\ldots,6 \}$. As a consequence, the number of independent kinematical variables is reduced to seven. In what follows, we always work in the frame $x_0 \to \infty$, which simplifies the conformal cross-ratios to ratios of Mandelstam variables $x_{ij}^2/x_{kl}^2$. 

\subsection{Leading singularities for Wilson loops with Lagrangian insertion}
\label{sec:leadingsingularities}

In~\cite{Chicherin:2022bov,Chicherin:2022zxo,Brown:2025plq}, general results for leading singularities of the $L$-loop $n$-point observable were discussed. 
In reference~\cite{Brown:2025plq}, the latter were proved to be linear combinations of so-called {\it Kermit functions}, whose definition in momentum twistors is
\begin{equation}
    B_{ijklm}:=\frac{\langle AB(mij)\cap(jkl)\rangle^2}{\langle ABjm\rangle\langle ABij\rangle\langle ABjk\rangle\langle ABlj\rangle\langle ABmi\rangle\langle ABkl\rangle} \,,
\end{equation}
with the shorthand notation $(abc)\cap(def):=(ab)\langle cdef\rangle-(ac)\langle bdef\rangle+(bc)\langle adef\rangle$, together with special case
\begin{equation}
    B_{ijkl}:=\frac{\langle ijkl\rangle^2}{\langle ABij\rangle\langle ABjk\rangle\langle ABkl\rangle\langle ABli\rangle}\,.
\end{equation}
Here we always have the cyclic constraint $1\leq i<j<k<l<m\leq n$, due to planarity. In the frame of $x_0\to\infty$, $(AB)$ is identified with $ I_\infty$, and we denote the Kermits by $b_{ijklm}$ and $b_{ijkl}$ respectively. 

According to this result, the number of independent leading singularities is equal to the number of independent Kermits. Naively, there are $\left(\substack{n\\5}\right)$ and $\left(\substack{n\\4}\right)$ for two types of Kermits, together with their cyclic images. 
However, at six points, only  
$20$ of these are linearly independent. 

On the other hand, choosing a basis of linearly-independent Kermits has a disadvantage -- it breaks the dihedral symmetry of the Wilson loop. For this reason, we find it convenient to work with an over-complete set of $22$ leading singularities, which enjoys simple dihedral relations. 
Explicitly, we make the following choice of rational functions, 
\begin{equation}\label{eq:LS6}
    \begin{aligned}
        &R_1 = b_{12346}{+}b_{12456}{+}b_{23456}{-}b_{1256}{-}b_{2356}{-}b_{3456}\,, \\
        &R_2 = b_{12346}{+}b_{12456}{-}b_{1256} \,, \\
        &R_8 =b_{12346}{+}b_{12456}{+}b_{23456}{+}b_{1234}{+}b_{1456}{-}b_{1256}{-}b_{2356}{-}b_{3456} \,,\\
     &R_{11} = b_{2356} \,,\\
        &R_{14} =b_{1234}{-}b_{1456} \,,\\  &R_{17} =b_{1246}{+}b_{1345} \,, \\ &R_{20}=b_{1246}{-}b_{1345} \,,
    \end{aligned}
\end{equation}
as well as the following cyclic images, $\{R_{2+i}\}_{i=1,\cdots,5}$, $\{R_{8{+}i},R_{11{+}i},R_{17{+}i},R_{20{+}i}\}_{i=1,2}$. The latter are obtained by acting with the translation $\tau(Z_i) = Z_{i+1}$ $i$ times. 
The two linear relations that reduce these $22$ rational functions to $20$ independent ones are as follows,
\begin{equation}\label{eq:ldLS}
    \begin{aligned}
0 &= R_2-R_3+R_4-R_5+R_6-R_7-R_{20}+R_{21}-R_{22}  \,,\\
0 &= 4 R_1-2 R_3-2 R_5-2 R_7+R_{17}+R_{18}+R_{19}-R_{20}+R_{21}-R_{22}  \,. 
\end{aligned}
\end{equation}
Let us discuss the properties of the leading singularities under the hexagonal dihedral group, consisting of
 translation $\tau$ and reflection $\rho$ (defined by $\rho(Z_i) = Z_{7-i}$). ($\rho\tau\rho\tau= \tau^6= \rho^2 = 1$.)
 The $22$ leading singularities $R_i$ are organized into five orbits, cf. Table \ref{Tab:1}. One can see that if we chose a set of 20 linearly-independent leading singularities, we would lose the simple transformation laws under $\tau$ and $\rho$, since eqs.~\eqref{eq:ldLS} lead to linear relations among terms on different orbits 
\begin{align}
\{1\} , \{2,\ldots,7\} , \{ 17,18,19\}, \{20,21,22\} \,.\label{orbld}
\end{align}
For this reason we prefer to work with the $22$ leading singularities and apply eqs.~\eqref{eq:ldLS} at a later stage.

\begin{table}[t]
\centering
\begin{tabular}{|c|c|c|c|c|}
\hline
label & length & orbit &  relations & $i$ of  $R_i, G_i^{(L)}$ \\ \hline\hline
$\#1$ & 1& $\{A \}$ & $A=\tau(A)=\rho(A)$  & $\{1\}$ \\ \hline
$\#2$  & 6 & $\{ A, \tau(A), \ldots, \tau^5(A) \}$ & $A = \rho \tau^5 (A)$ & $\{2,3,4,5,6,7\}$ \\\hline
$\#3$& 3 & $\{ A, \tau(A), \tau^2(A) \}$ &  $A=\tau^3(A) = \rho \tau^2 (A)$ & $\begin{matrix}\{ 8,9,10\},\\\{ 11,12,13\},\\\{ 17,18,19\}\end{matrix}$ \\\hline
$\#4$ & 3 & $\{ A, \tau(A), \tau^2(A) \}$ &  $A=-\tau^3(A) = \rho \tau^2 (A)$ & $\{ 14,15,16\}$ \\\hline
$\#5$ & 3 & $\{ A, \tau(A), \tau^2(A) \}$ &  $A=-\tau^3(A) = -\rho \tau^2 (A)$ & $\{ 20,21,22\}$ \\
\hline
\end{tabular}
\caption{Dihedral orbits of $22$ leading singularities}\label{Tab:1}
\end{table}

\subsection{Iterated integrals, differential equations and symbols}
In this work, integrated level results are always multi-polylogarithmic (MPL) functions, and specially linear combinations of Goncharov polylogarithms. The latter are defined recursively as~\cite{goncharov2005galois,Duhr:2019tlz}
\begin{equation}
G(a_1,\cdots,a_n,z):=\int_0^z\frac{{\rm d}t}{t-a_1}G(a_2,\cdots,a_{n};t)\,,
\end{equation}
with $G(;z):=1$, and $G(a_1,\cdots,a_n;z):=\frac{1}{n!}\log^n z$ if $(a_1,\cdots,a_n){=}(0,\cdots,0)$. An MPL function with $w$-fold integrations is said to have {\it weight} $w$, and logarithmic function is said to be weight-one. The total differential of a weight-$w$ MPL function $\mathcal{F}^{(w)}$ always has the following form
\begin{equation}
    {\rm d}\mathcal{F}^{(w)}=\sum_i\mathcal{F}_i^{(w{-}1)}{\rm d}\log x_i \ , 
\end{equation}
where $\mathcal{F}_i^{(w{-}1)}$ are weight-$(w{-}1)$ MPL functions, and $x_i$ are functions of $\mathcal{F}^{(w)}$'s variables. Following this structure, the {\it symbol} \cite{Goncharov:2010jf, Duhr:2011zq} of MPL function is iteratively defined as 
\begin{equation}
    \mathcal{S}(\mathcal{F}^{(w)})=\sum_i\mathcal{S}(\mathcal{F}_i^{(w{-}1)})\otimes x_i \ . 
\end{equation}
The symbol of $\mathcal{F}^{(w)}$ is written as a length-$w$ tensor product.
\begin{equation}
    \mathcal{S}(\mathcal{F}^{(w)})=\sum_I c_I\ a_1^{I}\otimes\cdots\otimes a_w^{I} \ .
\end{equation}
In general, the coefficients $c_I$ are also functions of the external kinematics. We  call these coefficients {\it leading singularities}, noting that in the context of Feynman integrals there is also a related meaning of this term \cite{Arkani-Hamed:2010pyv}. In special cases when the $c_I$'s are rational numbers, $\mathcal{F}^{(w)}$ is called a {\it pure function}. The entries $a_j^{I}$ in the product are called {\it symbol letters}. They are functions of kinematical variables. The set of symbol letters for an MPL function is called its {\it symbol alphabet}.   Describing physical singularities of $\mathcal{F}^{(w)}$, leading singularities and symbol letters provide a basic understanding of analytic structures of an MPL function.

In the differential equation approach \cite{Henn:2013pwa}, MPL functions and their symbol letters also play an important role. For a family of MPL master integrals in dimension regularization, after turning them into a uniform transcendental (UT) basis, they satisfy the canonical differential equations
\begin{equation}
    {\rm d}\mathbf{I}=\epsilon\ {\rm d}A\cdot\mathbf{I}\,,
\end{equation}
with regularization parameter $\epsilon$ factorized out. The matrix ${\rm d}A$ only depends on kinematical variables as
\begin{equation}
    {\rm d}A=\sum_{i}A_i\ {\rm d}\log W_i \,,
\end{equation}
where $A_i$ are number matrices, and $W_i$ are symbol letters of the master integrals. For the canonical differential equation calculation, prior knowledge of the symbol alphabet can avoid the complexity of reconstructing ${\rm d}\log(W)$-form in ${\rm d}A$ and turn calculations of canonical differential equations to be a linear fit problem for coefficient matrices $A_i$.

More importantly, if the symbol alphabet can be determined before integration, we can adopt the symbol bootstrap strategy to obtain amplitudes or integrals, instead of direct integration, which may turn out to be too complicated to perform when the number of scales increases. The basic idea of symbol bootstrap is that one can construct the function space and ansatz for the observable from symbol letters and integrability conditions. One can impose physical conditions the observable satisfies, get rid of the unnecessary candidates, and finally arrive at the result. Great progress has been made in bootstrapping scattering amplitudes of planar $\mathcal{N}=4$ sYM theory \cite{Dixon:2011pw,Dixon:2014xca,Dixon:2014iba,Drummond:2014ffa,Dixon:2015iva,Caron-Huot:2016owq,Dixon:2016nkn,Drummond:2018caf,Caron-Huot:2019vjl,Caron-Huot:2019bsq,Caron-Huot:2020bkp,Dixon:2023kop} following this idea, both at the symbol and function level, and also for form factors in this theory \cite{Dixon:2020bbt,Dixon:2021tdw,Dixon:2022rse,Dixon:2022xqh,Basso:2024hlx}.

\subsection{Basis of planar six-point two-loop functions}
\label{sec:symbols}

The relevant two-loop six-particle function space was recently computed in a series of works~\cite{Henn:2021cyv,Henn:2022ydo,Henn:2024ngj,Henn:2025xrc,Abreu:2024fei}.
Here we collect the relevant information at symbol level. 

We remark that in the usual symbol bootstrap approach one starts from a given symbol alphabet, and assembles integrable symbols up to a certain transcendental weight, as described in the previous subsection. In the present situation, thanks to the results of the above references, we can bypass this step and directly read off the symbols from the analytic results of the two-loop Feynman integrals computed there. A similar approach was taken in reference \cite{Guo:2021bym} for certain form factor integrals.
We also include symbols corresponding to products of one-loop integrals, as those are relevant to the Wilson loop observable. This information has been conveniently assembled in reference \cite{Henn:2025xrc}.

\begin{table}[t]
\centering
\begin{tabular}{|l|c|c|c|c|}
\hline
Transcendental weight  & 1 & 2 & 3 & 4 \\ \hline
\hline
\# All symbols  & 9 & 62 & 319 & 945 \\\hline
\# Two-loop six-point symbols  & 9 & 62 & 266 & 639 \\\hline
\# Two-loop five-point one-mass symbols   & 9 & 59 & 263 & 594 \\ \hline
\# One-loop squared symbols  & 9 & 59 & 221 & 428 \\ \hline
\# Genuine two-loop six-point symbols  & 0 & 0 & 3 & 45 \\ \hline
\end{tabular}
\caption{Counting of independent symbols for two-loop six-point massless planar Feynman integrals, cf. also reference \cite{Henn:2025xrc}.}
\label{tab:symbol_weights}
\end{table}

Up to the weight four (assuming external four-dimensional kinematics, and retaining terms up to and including the finite part only), these integrals contain $245$ symbol letters. 
In the notation of ref.~\cite{Henn:2025xrc}, they are a subset of $\{W_1, \ldots W_{289}\}$ which refer to the two-loop six-point alphabet with $D$-dimensional kinematics.
The letters are closed under dihedral symmetry. $156$ letters are parity even while $89$ are parity odd. Out of these letters, $232$ letters are from two-loop five-point integrals with one massive particle, while $13$ letters are genuine six-point letters,
\begin{equation}\label{eq:letter6}
\begin{aligned}
&W_{100}=-s_{23} s_{34} s_{56}+s_{23} s_{345} s_{56}-s_{12} s_{45} s_{61}-s_{34} s_{61} s_{123}+s_{12} s_{45} s_{234}+s_{34} s_{123} s_{234} \\
&\phantom{aaaaaaaaaaaaaaaaaaaaaaaaaaaaaaaaaaaaaaa}+s_{61} s_{123} s_{345}-s_{123} s_{234} s_{345}\,, \\
 &W_{100+i}=\mathcal \tau^i (W_{100}), \quad i=1,\ldots,5 \,, \\
&W_{138}=\Delta_6=
\langle12\rangle [23]\langle34\rangle [45]\langle 56\rangle [61]-[12]\langle 23\rangle [34]\langle45\rangle [56]\langle61\rangle \,, \\
&W_{242}=\frac{{-}s_{12} \left(s_{45}{+}s_{61}{-}s_{234}\right){+}s_{23} \left(s_{34}{+}s_{56}{-}s_{345}\right){+}s_{123} \left({-}s_{34}{+}s_{61}{-}s_{234}{+}s_{345}\right){-}\epsilon(1,2,3,5)}{{-}s_{12} \left(s_{45}{+}s_{61}{-}s_{234}\right){+}s_{23} \left(s_{34}{+}s_{56}{-}s_{345}\right){+}s_{123} \left({-}s_{34}{+}s_{61}{-}s_{234}{+}s_{345}\right){+}\epsilon(1,2,3,5)} \,,
\\
 &W_{242+i}=\mathcal \tau^i (W_{242}), \quad i=1,\ldots,5 \,,
\end{aligned}
\end{equation}
where $\tau$ stands for the cyclic permutation generator. Especially, families $\{W_{i{+}100}\}_{i=0,\cdots,5}$ and $\{W_{i{+}242}\}_{i=0,\cdots,5}$ arise from Feynman integral sectors in Fig~\ref{fig:PT}.

Table~\ref{tab:symbol_weights} shows the counting of symbols up to weight four \cite{Henn:2025xrc}.
 Note that two-loop five-point integrals with one massive particle are in the two-loop six-point massless planar Feynman integral families, thus their symbols span a linear subspace of six-point symbols. The dimension of the quotient space is the number of genuine two-loop six-point symbols. On the other hand, for the purpose of the bootstrap, in this paper we also need the symbols from the products of one-loop pure integrals. These symbols are listed in the ``one-loop squared symbols" row of Table~\ref{tab:symbol_weights}. Finally, the ``all symbols" row in this table refers to the union of two-loop six-point symbol space and the products of one-loop six-point symbols. This means that a given (pure) weight-four function in this space can be described by $945$ indeterminates only. We have not compared this against the number of integrable weight-four symbols that one would obtain from the $245$-letter alphabet (subject to first entry conditions), but it is likely that this number would be much larger.

To end this section, we offer an extra counting in Table~\ref{Tab:2} about numbers of functions that satisfy different dihedral behaviors according to Table~\ref{Tab:1}, {\it e.g.}, for the orbit $\#3$, we count the number of six-point planar functions $f_i$ at each weight satisfying $\tau^3(f_i)=\rho\tau^2(f_i)=f_i$, {\it etc.} This will be the starting point of our dihedral ansatz in the next section.

\begin{table}
\centering
\begin{tabular}{|l|c|c|c|c|c|}
\hline
orbit/weight&0 & 1 & 2 & 3 & 4 \\ \hline\hline
$\#1$&1 & 2 & 11 & 40 & 116 \\\hline
$\#2$&1 & 5 & 34 & 165 & 487 \\\hline
$\#3$&1 & 4 & 23 & 98 & 282 \\\hline
$\#4$&0 & 2 & 14 & 78 & 238 \\\hline
$\#5$&0 & 1 & 11 & 67 & 205 \\\hline
\end{tabular}
\caption{Number of functions with specific behavior under dihedral transformations, specified by the symmetry orbits given in Table~\ref{Tab:1}.}
\label{Tab:2}
\end{table}

\section{Bootstrapping the Wilson loop observable}
\label{sec: bootstrap and discussion}

Now we discuss the explicit bootstrap calculation. We start by reviewing the previous results at six-point tree and one-loop level, and then move to the two-loop case. We discuss the physical conditions and the corresponding constraints that we use in the bootstrap procedure. Afterwards, we show how the observable can be completely bootstrapped, and then we end with a discussion on the properties of the result.

\subsection{Previously known tree-level and one-loop results}

Let us present explicit results of lower-loop observables which were also discussed in \cite{Chicherin:2022bov}. At Born level, $F_6$ is just the one-loop six-particle MHV integrand
\begin{align}
F^{(0)}_6 = -R_1 \,. \label{eq:F60}
\end{align}
At the one-loop level, only nine leading singularities are present. We have
\begin{equation}\label{eq:F61}
    \begin{aligned}
F^{(1)}_6 = & R_2 \, {\rm Pent}_{2,6}  + R_3 \, {\rm Pent}_{1,3} + R_4 \, {\rm Pent}_{2,4} + R_5 \, {\rm Pent}_{3,5} + R_6 \, {\rm Pent}_{4,6} \\& + R_{7} \, {\rm Pent}_{1,5} 
 + R_8 \, {\rm Pent}_{1,4} + R_9 \, {\rm Pent}_{2,5} + R_{10} \, {\rm Pent}_{3,6} \,,
\end{aligned} 
\end{equation}
where ${\rm Pent}_{ij}$ is the chiral pentagon integral introduced in ref.~\cite{Arkani-Hamed:2010pyv}. It is given by 
\begin{align}
 {\rm Pent}_{ij}   &= \log u\log v+\text{Li}_2(1{-}u)+\text{Li}_2(1{-}v)+\text{Li}_2(1{-}w)-\text{Li}_2(1{-}uw)-\text{Li}_2(1{-}vw) \,,
\end{align}
where 
\begin{equation*}
u=\dfrac{x_{i{+}1,j{+}1}^2}{x_{i,j{+}1}^2} \,,\quad  v=\dfrac{x_{i,j}^2}{x_{i,j{+}1}^2} \,,\quad w=\dfrac{x_{i,j{+}1}^2x_{i{+}1,j}^2}{x_{i,j}^2x_{i{+}1,j{+}1}^2} \,.
\end{equation*}
The tree-level and one-loop formulas are in agreement with the bootstrap ansatz \eqref{eq:symbolbootstrapintroduction}. Indeed, the leading singularities appearing in those formulas form a subset of the basis discussed in subsection \ref{sec:leadingsingularities}, and the symbols of the pentagon integrals are part of the weight-two symbol space discussed in subsection \ref{sec:symbols}.
In the next subsection, we discuss how to recover those results via the symbol bootstrap, and then move on to bootstrap the new two-loop result.

\subsection{Symbol bootstrap for $F_6^{(2)}$}

Having discussed the lower-loop results, we now move to the actual bootstrap calculation for the observable. Our starting point is eq. \eqref{eq:symbolbootstrapintroduction} at six points.
As mentioned in subsection \ref{sec:leadingsingularities}, we find it convenient to start with an over-complete basis of $22$ leading singularities, as this simplifies the discussion of the dihedral symmetry. This leads to the concrete ansatz
\begin{align}
F_6^{(L)}= \sum_{i=1}^{22} R_i\, G^{(L)}_i \,, \label{eq:F6L}
\end{align}
where, in comparison with equation \eqref{eq:symbolbootstrapintroduction}, $R_i$ are the 22 leading singularities from subsection \ref{sec:leadingsingularities}, and 
\begin{align}
    G_i^{(L)}=\sum_{j}c_{ij}f_j^{(L)}\,
\end{align} are pure functions accompanying all leading singularities, as discussed in subsection \ref{sec:symbols}. 
More concretely, our ansatz at $L$ loops consists of symbols of uniform weight $2L$ \cite{Kotikov:2004er,Hannesdottir:2024hke}, which is a widely believed conjecture in $\mathcal{N}{=}4$ sYM theory.

It is also worth mentioning that due to the linear dependencies~\eqref{eq:ldLS} between the $R_i$, the number of unknown coefficients that are actually independent is
lower than that of the unknowns in our ansatz \eqref{eq:F6L}. This can be seen in Table~\ref{Tab:countingconstraints}. However, employing a redundant ansatz brings extra homogeneous freedom to the result, which can help us organize the result in a better manner, as we will show.

Our strategy is to impose the following constraints, in order to fix the unknowns in the ansatz:
\begin{enumerate}
    \item Dihedral symmetry of the observable.
    \item Overall scaling dimension.
    \item Cancellation of spurious singularities.
    \item Consistency with soft limit.
    \item Consistency with collinear limit.
    \item Consistency with triple collinear limit.
\end{enumerate}
We first construct a dihedral basis for each $G_i^{(2)}$ in \eqref{eq:F6L}, starting from  
the $945$ basis functions and dihedral behavior in Table~\ref{Tab:1}.
We then discuss the above constraints 2.-6. in more detail.
In principle one can impose those constraints in any order.
We find 
it interesting to count the number of independent constraints coming separately from each condition, cf. Table~\ref{Tab:countingconstraints}.
Of course, some of these constraints are not independent.
However, after combining all constraints together, we find that all independent unknowns are completely fixed.  
Specifically, the bootstrap procedure successfully reproduces the tree-level and one-loop results (taking weight zero and two symbols as input, respectively), and it gives a unique answer at two loops. We now discuss the constraints and their implementation in more detail.

\begin{table}[t]
\centering
\begin{tabular}{|l|c|c|c|c|c|}
\hline
weight&0 & 1 & 2 & 3 & 4 \\ \hline
\shortstack{{\bf unknowns} in \\ dihedral ansatz}&5 & 22 & 139 & 644 & 1892 \\
\hline
genuine {\bf unknowns}&4 & 20 & 125 & 585 & 1718 \\ \hline
{\bf constraints:}& & & & & \\
soft& 3  & 20 & 116 & 515 & 1439 \\
collinear& 3 & 20 & 121 & 551 & 1539 \\
spurious $s_{24} = 0$& 1 & 12 & 76& 360 & 1044 \\
spurious $s_{25} = 0$& 1 & 6 & 36 & 165 & 483 \\
scaling dimension&0 & 4 & 20 & 125 & 585 \\
triple collinear& 1 & 5 & 31 & 134 & 353 \\
\hline
total {\bf constraints}& 4 & 20 & 125 & 585 & 1718 \\ \hline
unfixed {\bf unknowns}&0 & 0 & 0 & 0 & 0 \\\hline
\end{tabular}
\caption{Numbers of constraints following from each physical condition.
}\label{Tab:countingconstraints}
\end{table}

{\it \underline{1. Dihedral symmetry.}}
We are looking for $F_6^{(2)}$ in the form of eq. \eqref{eq:F6L}, so we start by imposing the dihedral symmetries on the symbols. Following dihedral behavior of leading singularities $R_i$ in Table~\ref{Tab:1}, to make sure the ansatz~\eqref{eq:F6L} is also dihedral invariant, the function $G_i^{(L)}$ must have the same transformation law as $R_i$  for each $i$. The counting of symbols with given dihedral symmetry properties is given in Table~\ref{Tab:2}. 
Recall in subsection~\ref{sec:leadingsingularities} that $F_6^{(2)}$ involves one orbit of types $\#1,\#2,\#4,\#5$ and three orbits $\#3$ for the leading singularities. 
Thus, summing the numbers of symbols from these orbits, we find the number of unknowns in our ansatz \eqref{eq:F6L}, 
\begin{align}
 \text{\# of unknowns in ansatz} 
&=116+487+3\times282+238+205=1892\,.
\end{align}

{\it \underline{2. Scaling dimension.}}
The functions $G_{i}^{(2)}$ depend on dimensionless ratios of kinematical variables.
However, many of the symbol letters $W_i$ have a scaling dimension. (This is a natural choice in order to have simpler transformation properties under dihedral symmetry.) This means that some unknowns are fixed by requiring invariance under the rescaling $p_i \to \kappa p_i$.
This is a homogeneous condition on the ansatz. We find that it gives $585$ non-trivial conditions on the unknown coefficients at weight four.

{\it \underline{3. Cancellation of spurious singularities.}}
Some of the leading singularities have poles at $\langle 24\rangle \sim p_2 \cdot p_4 = 0$ and $\langle 25\rangle \sim p_2 \cdot p_5 = 0$. However, $F_6^{(2)}$ is finite in these limits. Therefore, these are  spurious singularities that have to be compensated by transcendental functions $G^{(L)}_{i}$.
This leads to homogeneous constraints on the ansatz.

To work out these constraints, we study which leading singularities become singular at $s_{24}=0$. We find the following residues at  $s_{24}=0$,
\begin{equation}
    \begin{aligned}
        R_4\to \tilde{R}_{17}{+}\tilde{R}_{18},\quad R_{13}\to\tilde{R}_{13},\quad R_{17},R_{20}\to\tilde{R}_{17},\quad R_{18}\to\tilde{R}_{18},\quad R_{21}\to{-}\tilde{R}_{18},
    \end{aligned}
\end{equation}
while the other $R_{i}$ yield vanishing residues. We denoted by $\tilde{R}_{i}$ (with $i=13,17,18$) three linearly-independent residues, whose precise expressions are unimportant. As a consequence, we obtain the following three constraints on the symbols,
\begin{align}
G_{4} + G_{17} + G_{20}|_{s_{24} = 0} = G_{4} + G_{18} - G_{21}|_{s_{24} = 0}=G_{13}|_{s_{24} = 0} = 0 \,.
\end{align}
Analogously, taking residues of the leading singularities at $s_{25}=0$, we find that only $R_{9}$ and $R_{15}$ have nonzero residues, and that the latter are linearly independent. Thus, the constraints on the transcendental functions are 
\begin{align}
G_{9}|_{s_{25} = 0} = G_{15}|_{s_{25} = 0} = 0 \,.
\end{align} 
These homogeneous conditions yield $1044$ and $483$ constraints on the unknowns, respectively.

{\it \underline{4. Consistency with soft limit.}}
We consider the soft limit $p_6 \to 0$. In this limit the light-like edge between $x_5$ and $x_6$ of the Wilson loop contour shrinks to a point, and the hexagonal contour becomes the pentagonal contour. We find that the $22$ leading singularities take the following forms in the soft limit
\begin{equation}
\begin{aligned}
	&R_1 \to -r_0 \, , \\
	&R_2 \to \bar{R}_2 \, , \quad &&R_{i=3,4,5} \to r_{i-1} - r_0 \, , \quad &&R_6 \to \bar{R}_6 \, , \quad R_7 \to 0 \\
	&R_8 \to r_5 - r_0 \, , \quad &&R_9 \to r_1 - r_0 \, , \quad &&R_{10} \to \bar{R}_{10} \, , \\
	&R_{11} \to \bar{R}_{11} \, , \quad &&R_{12} \to \bar{R}_{12} \, , \quad &&R_{13} \to r_3 \, ,  \\ 
    &R_{14} \to r_5 \, , \quad &&R_{15} \to r_1 \, , \quad &&R_{16} \to \bar{R}_{16} \, ,  \\ 
	&R_{17} \to \bar{R}_{17} \, , \quad &&R_{18} \to \bar{R}_{18} \, , \quad &&R_{19} \to R^\star \, , \\
	&R_{20} \to \bar{R}_{17} - 2r_2 \, , \quad &&R_{21} \to 2r_4 - \bar{R}_{18} \, , \quad &&R_{22} \to R^\star \, ,
\end{aligned}
\end{equation}
    where $R^\star{:=}\bar{R}_2{+}\bar{R}_6{-}\bar{R}_{17}{-}\bar{R}_{18}{+}r_0{+}r_{2}{+}r_3{+}r_4$, and where $r_i$ are the six leading singularities from the five-point observable, cf. eq. \eqref{eq:F52}. Finally, the precise form of $\bar{R}_i$ is again unimportant. What is important is that the set $\{ r_i , \bar{R}_i \}$ is linearly independent. Therefore the transcendental functions accompanying $ \bar{R}_i $ should vanish, and those accompanying $r_i$ should match the known five-point functions $g_i$, cf. \eqref{eq:F52}. Imposing these limits gives $14$ relations
\begin{equation}
    \begin{aligned}
        & G_1+G_3+G_4+G_5+G_8+G_9+G_{19}+G_{22}\to {-}g_{5,0}^{(2)} \,, \\
        & G_9+G_{15} \to g_{5,1}^{(2)} \,, \\
        &G_3+G_{19}-2G_{20}+G_{22}\to g_{5,2}^{(2)} \,, \\
        & G_4+G_{13}+G_{19}+G_{22}\to g_{5,3}^{(2)} \,, \\ 
        & G_5+G_{19}+2G_{21}+G_{22}\to g_{5,4}^{(2)} \,, \\
        &G_8+G_{14}\to g_{5,5}^{(2)} \,, \\
        &G_{i=10,11,12,16} \to0\,,\quad G_{17}-G_{19}+G_{20}-G_{22} \to0 \,,\\
        &G_{18}-G_{19}-G_{21}-G_{22} \to0\,,\quad G_ 2+G_{19}+G_{22}\to0 \,, \\
        &G_ 6+G_{19}+G_{22}\to0 \,.
\end{aligned}
\end{equation}
These yield $1439$ constraints on the unknown coefficients.

{\it \underline{5. Consistency with collinear limit.}}
We consider the collinear limit $p_6\to z\ p_5$, so the two edges from $x_4$ to $x_6$ become parallel and form a single light-like edge.
$F_6^{(2)}\to F_5^{(2)}$ under this limit. The $22$ leading singularities \eqref{eq:LS6} become 
\begin{equation}
    \begin{aligned}
            &R_1\to-r_0\,, \\
            &R_{i=2,3,4}\to r_{i-1}{-}r_0 \, , \quad &&R_5\to r_4{-}r_0\, , \quad  &&R_{i=6,7}\to0 \, ,\\
            &R_8\to r_5{-}r_0\, , \quad &&R_9\to r_1{-}r_0 \, , \quad &&R_{10}\to r_4{-}r_0 \,,\\
            &R_{11}\to 0 \, , \quad &&R_{12}\to r_2 \, , \quad &&R_{13}\to r_3 \,,\\
            &R_{14}\to r_5 \, , \quad &&R_{15}\to r_1 \, , \quad && R_{16}\to {-}r_4 \,,\\
            &R_{17}\to r_{2}{+}r_3 \, , \quad &&R_{18}\to r_4 \, , \quad &&R_{19}\to r_1 \,, \\
            &R_{20}\to r_3{-}r_2 \, , \quad &&R_{21}\to r_4 \, , \quad &&R_{22}\to r_1 \,.
    \end{aligned}
\end{equation}
The transcendental functions accompanying the $r_i$ from the five-cusp case should match the corresponding functions $g_{5,i}^{(2)}$, cf. \eqref{eq:F52}. Thus we get six relations at $p_6\to z\ p_5$,
\begin{equation}
    \begin{aligned}
        & G_1+G_2+G_3+G_4+G_5+G_8+G_9+G_{10}\to {-}g_{5,0}^{(2)}\, ,\\
        & G_2+G_9+G_{15}+G_{19}+G_{22} \to g_{5,1}^{(2)}\, ,\\
        &G_3+G_{12}+G_{17}-G_{20}\to g_{5,2}^{(2)}\, ,\\
        & G_4+G_{13}+G_{17}+G_{20}\to g_{5,3}^{(2)}\, , \\ 
        & G_5+G_{10}-G_{16}+G_{18}+G_{21}\to g_{5,4}^{(2)}\, , \\
        &G_8+G_{14}\to g_{5,5}^{(2)}  \, ,
\end{aligned}
\end{equation}
which yield 1539 constraints on the unknown coefficients.

{\it \underline{6. Consistency with triple collinear limit.}}
We let $p_4$, $p_5$, $p_6$ approach fractions of the light-like momentum $P$ ($P^2 = 0 $) at the same speed,
\begin{align}
p_4 \to z_1 P \,,\quad
p_5 \to z_2 P \,,\quad
p_6 \to z_3 P \,,\quad
\end{align}
where the positive coefficients $z_1,z_2,z_3$ satisfy $z_1 + z_2 + z_3 = 1$, and  $p_1 + p_2 +p_3 + P = 0$ in the limit. The observable reduces to the four-point one, $F_6^{(2)} \to F_4^{(2)}$, cf. \eqref{eq:formalF4} and \eqref{eq:F4}. Looking at the six-point leading singularities in the limit, they behave as
\begin{equation}
    \begin{aligned}
         R_1, R_2, R_3, R_4, R_9, R_{16} \to b_{1234} \,, \qquad 
         R_{14}, R_{18}, R_{21} \to {-}b_{1234}  \,,\\
\end{aligned}
\end{equation}
while all the remaining $R_i$ go to zero. Comparing to \eqref{eq:formalF4} and \eqref{eq:F4}, we get the following relation,
\begin{equation}
    G_1+G_2+G_3+G_4+G_9-G_{14}+G_{16}-G_{18}-G_{21}\to g_{4,0}^{(2)} \, ,
\end{equation}
which yields $353$ non-trivial constraints on the unknown coefficients.

{\it \underline{Summary:}}
We summarize all physical conditions we have considered in Table \ref{Tab:countingconstraints}. Weight-zero, -two and -four ansatz correspond to tree, one-loop and two-loop bootstrap respectively. For the sake of counting, we also include unphysical weight-one and -three bootstrap by requiring their soft,  collinear and triple collinear limits to be zero.  Line 2 and line 3 count numbers of unknown coefficients in all cases. Number of genuine unknowns is always smaller than number of unknowns in the dihedral ansatz at each weight, since there are two linear dependencies \eqref{eq:ldLS} among 22 leading singularities of our ansatz. From line four to line nine of the table we count numbers of independent constraints coming from the soft and collinear limits, absence of spurious singularities, scaling-dimensionless condition and triple collinear limit of the symbol. Line ten counts the total number of independent constraints, which match the number of genuine unknowns, and in this way all unknowns in the ansatz from weight zero to weight four are completely fixed.

\subsection{Discussion of the final symbol result}

As we have mentioned, there are redundancies in the dihedral ansatz \eqref{eq:F6L}, and the result is determined after we further expand the ansatz by $20$ independent leading singularities. On the other hand, if~\eqref{eq:ldLS} is not considered, extra unknowns still show up as in Table~\ref{Tab:countingconstraints}, whose accompanying functions are {\it homogeneous solutions} to our bootstrap problem, {\it i.e.}, they satisfy the desired dihedral structure, are dual conformal invariant and free of spurious poles, vanish under any considered physical limits, and vanish on the support of leading singularity relations \eqref{eq:ldLS}. If we keep the answer as~\eqref{eq:F6L} to retain the dihedral structure, we have $1892{-}1718=174$ homogeneous degrees of freedom to present our result. Two further constraints look natural for our organization. Firstly, the $174$ homogeneous solutions contain spurious letters, which only show up in these extra functions and drop out once we employ~\eqref{eq:ldLS}. Therefore, it would be better if we find a linear combination of homogeneous solutions which is free of these spurious letters. In this way, we can fix extra 100 unknowns and only 74 arbitrary coefficients are left. Secondly, according to the second relation in~\eqref{eq:ldLS}, we can use homogeneous solutions to eliminate the leading singularity $R_1$, which will furthermore reduce the number to $26$. Since $R_1$ is invariant under any dihedral transformation, such a fixing keeps the result in a dihedral structure. Finally, we do not have a preferred choice for the remaining $26$ coefficients, so we simply set them to zero.

We provide our main bootstrap result in ancillary files.
For convenience, these contain the basic definitions of the six-point alphabet and the leading singularities $R_i$. We also record two representations of the observable that were mentioned, including the expression with $21$ leading singularities (with $R_1$ eliminated), as well as the expression with $20$ linearly-independent leading singularities (without $R_{19}$ and $R_{22}$). All these results are represented in terms of symbols.

\begin{figure}[t]
    \centering
    \subfigure[$\{W_{i{+}10}\}_{i=0,\cdots,5}$]{\label{fig:DT1}\includegraphics[width=0.3\linewidth]{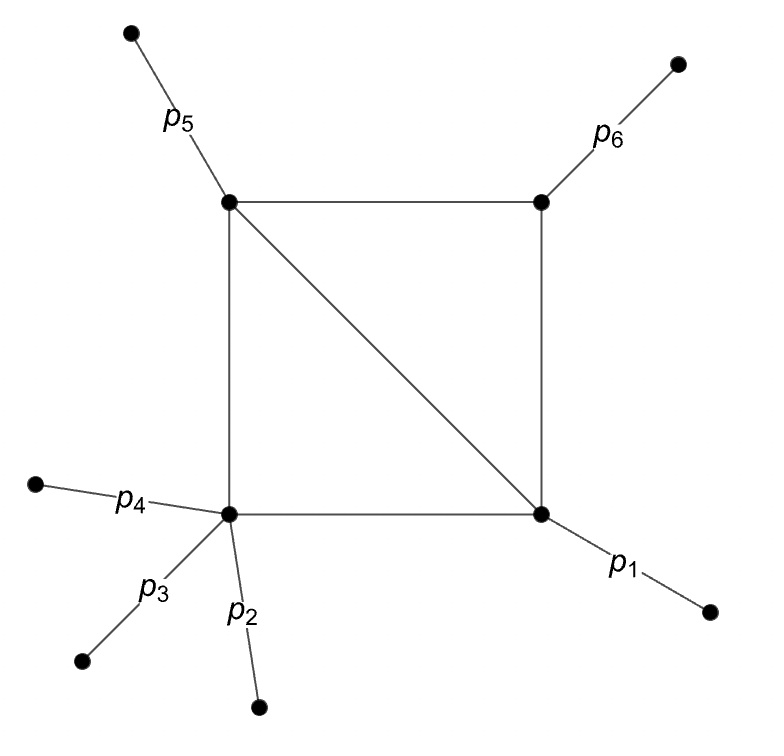}}
    \subfigure[$\{W_{i{+}34}\}_{i=0,\cdots,11}$]{\label{fig:DT2}\includegraphics[width=0.3\linewidth]{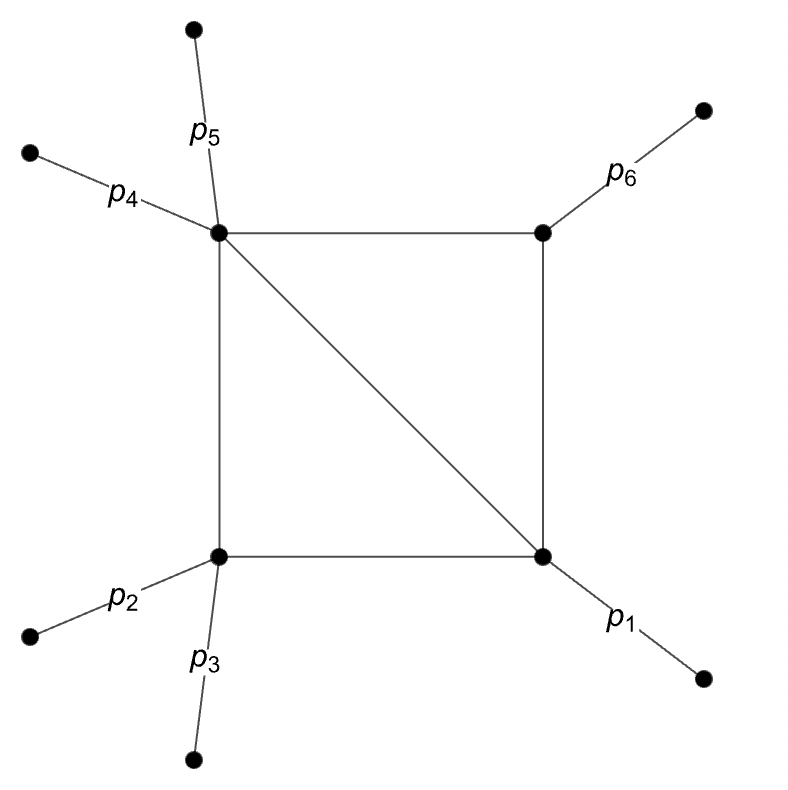}}
    \subfigure[$\{W_{i{+}58}\}_{i=0,\cdots,11}$]{\label{fig:BT}\includegraphics[width=0.35\linewidth]{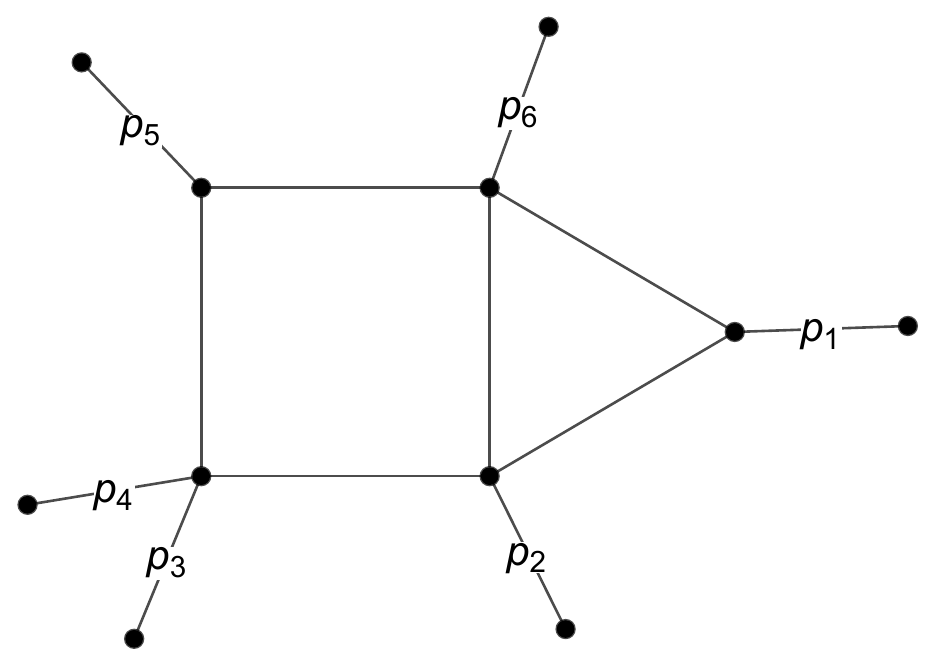}}
    \subfigure[$\{W_{i{+}76}\}_{i=0,\cdots,11}$]{\label{fig:DB}\includegraphics[width=0.35\linewidth]{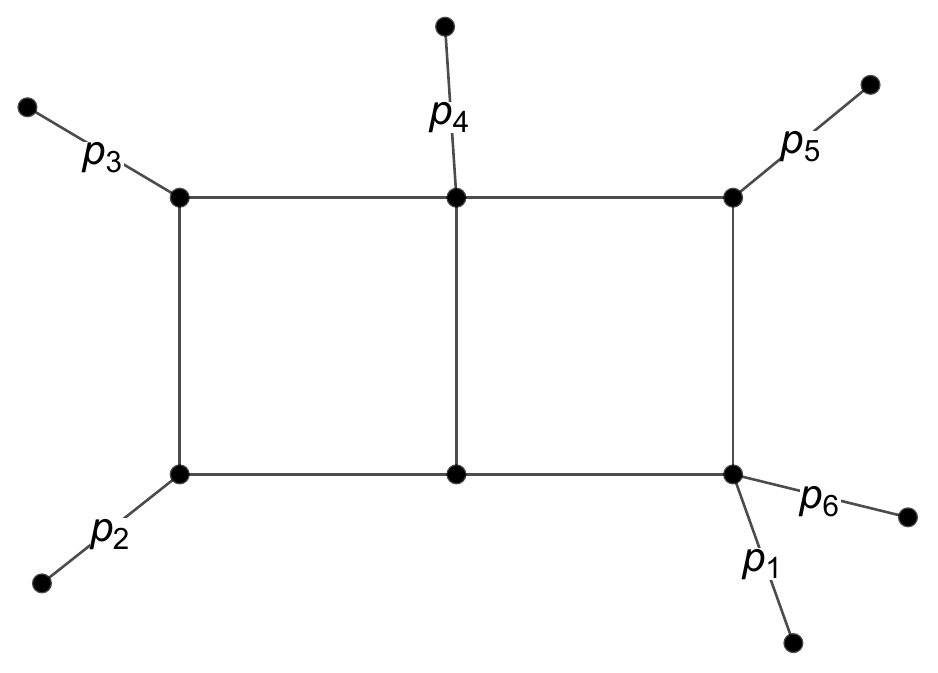}}
    \subfigure[$\{W_{i{+}100}\}_{i=0,\cdots,5}$]{\label{fig:PT}\includegraphics[width=0.35\linewidth]{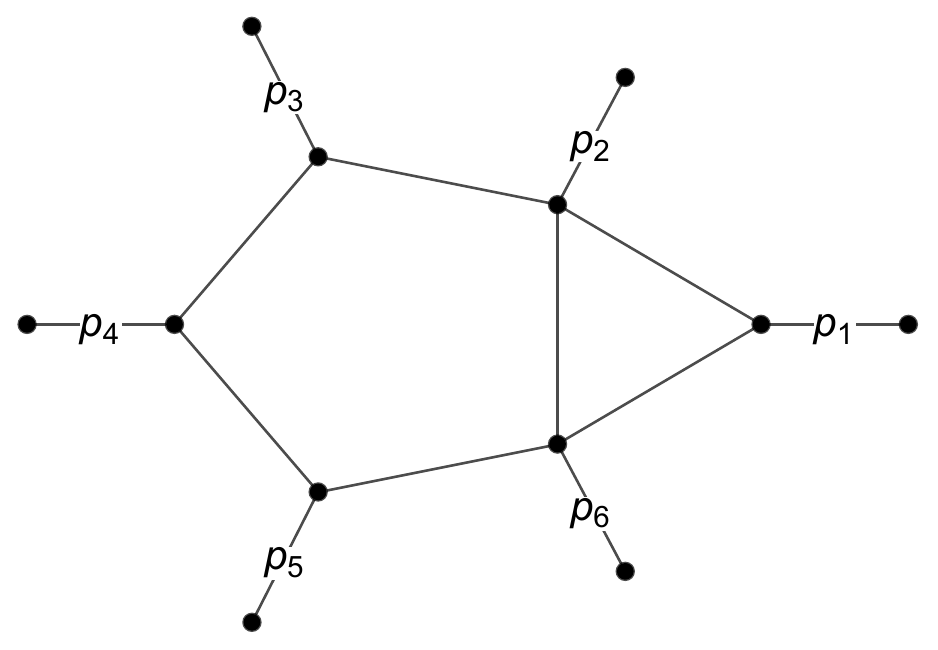}}
    \caption{Six-point planar Feynman integral sectors corresponding to genuine two-loop singularities that we see in the observable. 
    }
    \label{fig:FD6pt}
\end{figure}

Let us discuss the alphabet letters that appear in the final answer for $F^{(2)}_{6}$. While the full six-point two-loop alphabet consists of $245$ symbol letters, only $137$ of these appear in our bootstrap result for $F^{(2)}_{6}$.
These $137$ letters can be split into $87$ parity-even letters and $50$ parity-odd letters.
Let us discuss them in turn, using the notation of ref. \cite{Henn:2025xrc}.

$39$ of the even letters appeared already in one-loop six-point differential equations, cf. \cite{Henn:2022ydo}.
They are given by
   \begin{align}
   \{W_1,\cdots, W_9,W_{16},\cdots,W_{33},W_{46},\cdots,W_{51},W_{88},\cdots,W_{93}\} \,.
   \end{align} 
The remaining $48$ even letters are genuine two-loop letters, cf. \cite{Henn:2025xrc}. They can be grouped into five dihedral orbits as follows,
    \begin{align}
        \{W_{10{+}i}\}_{i=0,\cdots, 5}\ \cup \{W_{34{+}i}&\}_{i=0,\cdots, 11}\ \cup \{W_{58{+}i}\}_{i=0,\cdots, 11}\ \nonumber\\
        &\cup \{W_{76{+}i}\}_{i=0,\cdots, 11}\ \cup \{W_{100{+}i}\}_{i=0,\cdots, 5} \,. \label{eq:oddfrom}
    \end{align}
        These letters can be seen to come from the five two-loop Feynman integral sectors (and their dihedral images) shown in Figure \ref{fig:FD6pt}. Each of these letters can be associated with the leading Landau singularity of a specific UT two-loop master integral, and consequently they appear as diagonal elements in the canonical differential equation with respect to these master integrals \cite{He:2023umf}.  More precisely, we have the correspondence 
\begin{align}
    \{\mathcal{I}_{24}^{(pt)},\mathcal{I}_{19}^{(pt)},\mathcal{I}_{10}^{(pt)},\mathcal{I}_{30}^{(pb)},\mathcal{I}_{1}^{(pt)}\}\longleftrightarrow\{W_{13},W_{36},W_{62},W_{77},W_{100}\}\,.
    \end{align}

Let us now turn to the parity-odd letters. $36$ of them involve $\epsilon(i,j,k,l)$, namely   
    \begin{align}
    \{W_{182},\cdots,W_{190},W_{194},\cdots, W_{211},W_{218},\cdots,W_{220},W_{242},\cdots,W_{247} \}\,.
    \end{align}
Taking products of their numerators and denominators gives us parity-even expressions. It is interesting to know which singular locus the latter correspond to. We find that
$18$ of the letters correspond to one-loop singularities only, while the other $18$ encode two-loop singularities (specifically ones related to the last two dihedral families in \eqref{eq:oddfrom}). 
Finally, there are $7+7$ odd letters that involve 
the three-mass-triangle roots 
\begin{align}
r_1{=}&\sqrt{\lambda(s_{12},s_{34},s_{56})} \,,\qquad r_2{=}\sqrt{\lambda(s_{23},s_{45},s_{16})}\,,
\end{align}
where $\lambda(a,b,c):=a^2{+}b^2 {+}c^2{-}2ab{-}2ac{-} 2bc$. 
Interestingly, these letters appeared already at one loop (cf. \cite{Henn:2022ydo}). 
They read
    \begin{align} \label{eq:nonrational}
        \{W_{157},\cdots,W_{166},W_{275},\cdots,W_{278}\} \,.
            \end{align}
 Note that last four of these odd letters contain 
 $r_i$ and $\epsilon(i,j,k,l)$ simultaneously.

Let us make the following comments. All genuine six-point letters, cf. eq. ~\eqref{eq:letter6}, except for $W_{138}=\Delta_6$, appear in the final result.
Note that of the $137$ letters, $123$ are rationalized if we employ momentum twistors or spinor-helicity variables, and only the $14$ letters of eq. (\ref{eq:nonrational}) involve square roots.

It is worth mentioning that, in the final result, the functions accompanying $R_{11}, R_{12}$ and $R_{13}$ are particularly simple: they can be expressed in terms of products of one-loop chiral pentagons as follows,
\begin{equation}
    G_{11}^{(2)}=\text{Pent}_{2,6}\times\text{Pent}_{3,5},\ G_{12}^{(2)}=\text{Pent}_{1,3}\times\text{Pent}_{4,6},\ G_{13}^{(2)}=\text{Pent}_{1,5}\times\text{Pent}_{2,4} \,.
\end{equation}
It would be interesting to find a simple explanation of this observation, and perhaps an all-$n$ generalization of it. 

Our symbol result captures the full dependence on seven dimensionless variables in the four-dimensional on-shell six-particle kinematics. In the process of bootstrapping this result, we have already explored various important physical limits. However, there may be other limits of interest. Here we restrict ourselves to mentioning Appendix \ref{app:remarks}, which explores the multi-Regge limit.
We leave a more detailed analysis, possibly also including beyond-the-symbol terms, or other interesting limits, to future work.

\subsection{Duality to all plus YM amplitudes and Steinmann relations}

Besides the Wilson loop/scattering amplitude duality that exists within $\mathcal{N}=4$ sYM theory, Wilson loops with a Lagrangian insertion have also been observed \cite{Chicherin:2022bov,Chicherin:2022zxo} to be dual to the maximal transcendental part of amplitudes in pure YM theory where all gluons have the same helicity. This duality is expressed in the following relation,
\begin{equation}
     \vev{ W_n \mathcal{L}(x_0)}_{x_0 \rightarrow \infty} \sim \frac{\mathcal{A}_n^{\text{YM}}(1^+,\dots,n^+)}{\mathcal{A}_n^{\text{YM},(1)}(1^+,\dots,n^+)} \, , 
\end{equation}
where the RHS is divided by the rational one-loop all-plus amplitude $\mathcal{A}_n^{\text{YM},(1)}$ to ensure that the total helicity weight is zero. Moreover, the $\sim$ sign indicates that the relation is established for the leading transcendental parts of both sides in the planar limit, modulo scheme differences. These scheme differences can be made explicit by writing the all-plus amplitude as a product of two terms as follows,
\begin{equation}
    \frac{\mathcal{A}_n^{\text{YM}}}{\mathcal{A}_n^{\text{YM},(1)}} = \mathcal{Z}_{\text{IR}}^{\text{YM}}(\epsilon) \mathcal{H}^{\text{YM}}_n \, ,
\end{equation}
where the first term encompasses the IR-subtraction scheme, while the second is the finite remainder function. As explained previously, Wilson loops are also dual to MHV scattering amplitudes in $\mathcal{N}=4$ sYM theory, which is translated in the relation
\begin{equation}
       \vev{ W_n }\sim \frac{\mathcal{A}^{\text{MHV}}_n}{\mathcal{A}^{\text{MHV,tree}}_n} = \mathcal{Z}^{\text{MHV}}_{\text{IR}}(\epsilon) \mathcal{H}^{\text{MHV}}_n \, ,
\end{equation}
where, once again, we have decomposed the MHV amplitude into a product of an IR-scheme factor $\mathcal{Z}$ with a finite remainder function $\mathcal{H}^{\text{MHV}}$. Since the duality of all-plus amplitudes is established exclusively at the level of maximal transcendental weight, then, by virtue of the maximal transcendentality principle, we can simply set the subtraction schemes to be equal, as well as being a pure function of two-particle invariants $s_{i,i+1}$. This implies that the subtracted remainder functions $\mathcal{H}$ satisfy the Steinmann relations. Hence, combining both of the duality statements above, we can establish a relation between observable $F_n$ and the two remainder functions at leading color in the maximal-transcendental weight, 
\begin{equation}
    F_n \mathcal{H}^{\text{MHV}} \sim \mathcal{H}^{\text{YM}} \, .
\end{equation}
Furthermore, we can expand each contribution to the duality above in the coupling in order to get an order-by-order statement of the duality. For the observable $F_6$, the LHS can be expanded as
\begin{equation}
    F_6 \times {\cal H}_6 = g^2\, F^{(0)}_6 + g^4 \left( F^{(1)}_6 + F^{(0)}_6 \, \mathcal{H}^{(1)}_6 \right) + g^6 \left( F^{(2)}_6 + F^{(1)}_6 \mathcal{H}^{(1)}_6 + F^{(0)}_6 \mathcal{H}^{(2)}_6 \right) + O(g^8)\, ,
\end{equation}
which, at each order in the perturbative expansion, coincides with the maximal transcendental piece of $\mathcal{H}^{\text{YM,}(L)}$. Combining this with the fact that the $\mathcal{H}^{\text{MHV},(L)}$ satisfy the Steinmann relations by themselves, we expect the following constraint at two loops
\begin{equation}\label{eq:Steinmann}
    \underset{s_{i,i+1,i+2} = 0}{\rm Disc}\; \underset{s_{i-1 , i , i+1} = 0}{\rm Disc} \left(  F^{(2)}_6 + F^{(1)}_6 \mathcal{H}^{(1)}_6 \right) = 0\,.
\end{equation}
Indeed, we can easily verify that our bootstrapped result for $F_6$ satisfies \eqref{eq:Steinmann}. 
Furthermore, we have also verified the conjecture made in \cite{Caron-Huot:2019bsq,Caron-Huot:2020bkp} regarding extended Steinmann relations. In particular, we found that, for any non-negative integer $k$, any $k$-fold discontinuity of the physical combination $F^{(2)}_6 + F^{(1)}_6 \mathcal{H}^{(1)}_6$ also satisfies Steinmann relations.

\section{Summary and Outlook}
\label{sec: outlook}
In this work, we computed hexagonal Wilson loop with a Lagrangian insertion in planar $\mathcal{N}=4$ sYM theory at two-loop order by six-point two-loop planar massless function space and symbol bootstrap approach. Started from a redundant ansatz containing 22 leading singularities, where each leading singularity was expected to be accompanied by a weight-four multi-polylogarithmic function, we constructed our functional basis based on 945 two-loop planar hexagon functions from recent canonical differential equation result. Furthermore, through imposing physical conditions as dihedral symmetry, dual conformal invariance, special physical limits {\it etc.}, the observable was fixed at the symbol level. As an analog for calculating scattering amplitude from symbol bootstraps at six points, our calculation shared a similar manner with the former procedure, but the result we got enjoys a richer function space and singularity structure. More importantly, our result is the first physical observable that lives in the new two-loop six-point planar function space, and sheds new light on the frontier of QCD perturbative calculations. 

Besides having provided new insights, our work also uncovers the following critical questions for future research.

{\it \underline{1. Extension of the result to function level.}}\\
In this work, we explored the observable at symbol level only. 
This means that certain beyond-the-symbol terms are currently not included. While these are not expected to fundamentally change the structure of the result, they are required for numerical evaluation, for example. 
It is therefore natural to extend this analysis to function level.
A natural possibility of achieving this would be to extend our bootstrap analysis to Chen iterated integrals.
Implementing the various physical constraints requires expanding those integrals in various limits, which can be conveniently achieved in the differential equation formalism, cf. for example \cite{Caron-Huot:2020vlo, Henn:2024qjq}.  
Another possibility, avoiding the bootstrap, would be to perform
an integration-by-part reduction onto the two-loop planar Feynman integrals. Such an approach was followed for example for the five-point observable at two loops \cite{Chicherin:2022zxo} and for individual negative geometries \cite{Chicherin:2024hes}.
Obtaining the result at function level would allow one to explore further physical properties of the observable.
This would be particularly interesting in view of testing conjectured positivity \cite{Chicherin:2022zxo} and complete monotonicity property \cite{Henn:2024qwe} of the answer.

{\it \underline{2. Special physical limits of the observable.}}\\
 Given the new result presented in this work, it is now possible to further study different kinematical limits that would not be accessible for fewer number of particles.
 In particular, one can study the multi-Regge limits~\cite{Regge:1959mz,DelDuca:2022skz}. We have found an interesting simplification at the symbol level, reported on in Appendix~\ref{app:remarks}. It would be interesting if this could be explained from first principles.
 Furthermore, another interesting limit that was considered previously in the context of scattering amplitudes in sYM is the self-crossing limit ~\cite{Dorn:2011gf,Dorn:2011ec,Dixon:2016epj}. 
     Given  the discussion in reference \cite{Dixon:2016epj}, it would be particularly interesting to supplement our result with an extra analysis on how to perform the appropriate analytic continuation  needed to probe this limit.

{\it \underline{3. Cluster algebraic structure of the two-loop six-point alphabet.}}\\
In $\mathcal{N}=4$ sYM theory, symbol alphabets for scattering amplitudes were discovered to enjoy a {\it cluster algebraic structure}.
At $n=6,7$ points, this is related to the Grassmannian ${\rm Gr}(4,n)$ \cite{Golden:2013xva}, while at $n=8$ \cite{Caron-Huot:2011zgw,Zhang:2019vnm} the tropical Grassmannian ${\rm Gr}(4,8)$ \cite{Henke:2019hve,Drummond:2019qjk,Arkani-Hamed:2019rds,Early:2023tjj} plays a crucial role. 
Related mathematical structures were also found beyond sYM theory, namely in individual Feynman integrals \cite{Caron-Huot:2018dsv,Henn:2018cdp,Chicherin:2020umh,He:2021esx,He:2021non,Aliaj:2024zgp},
as well as for the pentagon function alphabet to two loops. 
The latter case involves a generalization of ${\rm Gr}(3,6)$ corresponding to {\it flag varieties}~\cite{Bossinger:2022eiy,Bossinger:2024apm}. The function space we operate in corresponds to the six-particle generalization.
Given the above background, it is exciting to explore whether the symbol we found in this work has any special properties, such as relations to cluster algebras or tropical geometry. 
It is particularly interesting whether the $137$ alphabet letters that appear play any special role from this viewpoint. Related to this, it would be interesting if the Steinmann relations we observed in this work could be given some cluster algebra interpretation.

\section*{Acknowledgments}
We thank Elia Mazzucchelli for feedback on the manuscript.
This work was supported by the European Union (ERC, UNIVERSE PLUS, 101118787). Views and opinions expressed are however those of the authors only and do not necessarily reflect those of the European Union or the European Research Council Executive Agency. Neither the European Union nor the granting authority can be held responsible for them. D.C. is supported by ANR-24-CE31-7996.  Y.Z. is supported from the NSF of China through Grant No. 12247103.

\appendix

\section{Previous four- and five-point results}
\label{app:fourfive}

In this Appendix, we summarize present results at four and five points
\cite{Alday:2013ip,Chicherin:2022bov,Chicherin:2022zxo} that serve as boundary conditions in the bootstrap when we consider soft/(multi-)collinear limits.

The expressions for $F_4$ involve one independent leading singularity only, 
\begin{equation}\label{eq:formalF4}
    F_4^{(L)}=b_{1234}\ g_{4,0}^{(L)}\,.
\end{equation}
$b_{1234}$ can also be written as $st$, with $s=(p_1+p_2)^2$ and $t=(p_2+p_3)^2$. The first three orders of $g_{4,0}^{(L)}$ read~\cite{Alday:2013ip} 
\begin{align}
    &g_{4,0}^{(0)}=-1 \,, \\
    & g_{4,0}^{(1)}=\log^2(z){+}\pi^2 \,,   \\
    &g_{4,0}^{(2)}={-}\frac12\log^4(z){+}\log^2(z)\left[\frac23\text{Li}_2\left(\frac{1}{z{+}1}\right){+}\frac23\text{Li}_2\left(\frac{z}{z{+}1}\right){-}\frac{19\pi^2}{9}\right]\label{eq:F4}\\
    &{+}\log(z)\left[4\text{Li}_3\left(\frac{1}{z{+}1}\right){-}4\text{Li}_3\left(\frac{z}{z{+}1}\right)\right]{+}\frac23\left[\text{Li}_2\left(\frac{1}{z{+}1}\right){+}\text{Li}_2\left(\frac{z}{z{+}1}\right){-}\frac{\pi^2}6\right]^2\nonumber\\
    &{+}\frac{8\pi^2}3\left[\text{Li}_2\left(\frac{1}{z{+}1}\right){+}\text{Li}_2\left(\frac{z}{z{+}1}\right){-}\frac{\pi^2}6\right]{+}8\left[\text{Li}_4\left(\frac{1}{z{+}1}\right){+}\text{Li}_4\left(\frac{z}{z{+}1}\right)\right]{-}\frac{23\pi^2}{18} \,, \nonumber
\end{align}
where 
\begin{align}
    z=\frac{x_{1,3}^2}{x_{2,4}^2}=\frac{s}t \,.
\end{align}
For $F_5$, we have six linearly-independent $r_i$, namely
\begin{equation}
    r_0=b_{1245}{+}b_{2345}{-}b_{12345}\,,\qquad r_1=b_{2345}
\end{equation}
and $\{r_{1{+}i}\}_{i=1,\cdots,4}$ from cyclic images of $r_1$. At Born level,
\begin{align}
F_5^{(0)} = r_0\,.
\end{align}
At one-loop level, five linearly-independent leading singularities contribute. Starting from two-loop level, all six leading singularities contribute, and we have 
\begin{align}\label{eq:F52}
&F_5^{(1)} = \sum_{i=1}^{5} (r_i-r_0)\, \text{Pent}_{i{-}1,i{+}1} \,,\\
&F_5^{(2)}=\sum_{i=0}^5 r_i\, g_{5,i}^{(2)} \,.
\end{align}
In the five-point case, the dihedral properties of the leading singularities are much simpler. $r_0$ is invariant, and $r_1,\ldots,r_5$ belong to one cyclic orbit. For reflection $\rho(Z_i)\to Z_{6{-}i}$, the leading singularities (and also corresponding $g_{i,5}^{(L)}$) behave as
\begin{align}
\rho_5(r_0)=r_0 \,,\quad
\rho_5(r_i) = r_{6-i} \,,\quad i=1,\ldots,5 \,,
\end{align} 
Explicit results for $g_{5,i}^{(2)}$ were computed in~\cite{Chicherin:2022bov,Chicherin:2022zxo} in terms of the pentagon functions \cite{Chicherin:2017dob,Gehrmann:2018yef}.

\section{Multi-Regge limit behavior}
\label{app:remarks}

We finally study the observable in the multi-Regge kinematics \cite{Regge:1959mz,DelDuca:2022skz}. We consider the channel $56 \to 1234$. The transverse momenta are comparable 
\begin{align}
|{\bf p}_1| \simeq |{\bf p}_2| \simeq |{\bf p}_3| \simeq |{\bf p}_4|
\end{align}
but rapidities are strongly ordered,
\begin{align}
|p^{+}_1| \gg |p^{+}_2| \gg |p^{+}_3| \gg |p^{+}_4| \,, \\
|p^{-}_1| \ll |p^{-}_2| \ll |p^{-}_3| \ll |p^{-}_4| \,,
\end{align}
where $p_i = (p^{+}_{i},p^{-}_{i},{\bf p}_i)$. In practice, we find it more convenient to use the following parametrization (the following expression keeps only the leading term at $x \to 0$)
\begin{align}
& s_{12} = \frac{s_1}{x} \,,\quad s_{23} = \frac{s_2}{x} \,,\quad s_{34} = \frac{s_3}{x} \,,\quad s_{56} = \frac{s_1 s_2 s_3}{\kappa^2 |z_1-z_2|^2 x^3}  \,,\quad s_{345} = -|z_2|^2 \kappa \notag\\
& s_{123} = \frac{s_1 s_2}{\kappa  |z_1-z_2|^2 x^2} \,,\quad s_{234} = \frac{s_2 s_3}{\kappa x^2} \,,\quad s_{16} = - |z_1|^2 \kappa  \,,\quad s_{45} = -|1-z_2|^2\kappa
\end{align}
where $s_1,s_2,s_3,\kappa$ are dimensionful and $z_1,\bar{z}_1,z_2,\bar{z}_2$ are dimensionless, and we denote $|z_i|^2 = z_i \bar{z}_i$. In the Euclidean region, $z_i,\bar{z}_i$ are independent variables. At tree-level at $x \to 0$,
\begin{align}
F^{(0)}_6 = \frac{s_1 s_2 s_3 \bar{z}_1 (1-z_2)}{\kappa |z_1 - z_2|^2 x^3} + O(1/x^2) \,,
\end{align}
and the 22 leading singularities simplify a lot at $x \to 0$, namely 
\begin{align}
R_1, R_2 , R_5 , R_6, R_7,  R_9,  R_{10} ,  R_{14} = -F^{(0)}_6 + O(1/x^2) \,.
\end{align}
The remaining leading singularities are sub-leading in the limit. So we find at general $L$, 
\begin{align}
\frac{F^{(L)}_6}{F^{(0)}_6} = -G^{(L)}_1-G^{(L)}_2-G^{(L)}_5-G^{(L)}_6-G^{(L)}_7-G^{(L)}_9-G^{(L)}_{10}-G^{(L)}_{14} +O\left(\log^{2L}(x)/x^2\right) .
\end{align} 
At one-loop level, this limit reads
\begin{align}
& \frac{F^{(1)}_6}{F^{(0)}_6} = - 9 \log^2(x) + 2 \log(x)\log \left(\frac{s_1^3 s_2^3 s_3^3}{\kappa^9 |z_1 z_2(1-z_2)|^2
 |z_1-z_2|^3}\right) -\log ^2\left(\frac{s_1 s_2 s_3}{\kappa ^3}\right) \notag\\ 
&  +2\log \left(|z_1(z_1-z_2)|^2\right) \log \left(\frac{s_1}{\kappa }\right)+2 \log \left(|z_2(z_1-z_2)|^2\right)
   \log \left(\frac{s_2}{\kappa }\right)  \notag\\ 
&  +2 \log
   \left(|(z_1-z_2)(1-z_2)|^2\right) \log
   \left(\frac{s_3}{\kappa }\right)+\log \left( |z_2|^2\right) \log
   \left(\frac{|z_2|^2}{|z_1(1-z_2)|^2}\right)  \notag\\ 
&   -\log
   \left(|z_1-z_2|^2\right) \log \left(|z_1 z_2 (z_1-z_2)|^2\right) + O(x \log^2(x))\,
\end{align}
modulo transcendental constants like $\pi^2$. At two-loop level, we observe a remarkable simplification at the symbol level,
\begin{align}
 \frac{F^{(2)}_6}{F^{(0)}_6} - \frac{1}{2}\left( \frac{F^{(1)}_6}{F^{(0)}_6} \right)^2  \sim O(x \log^4(x)) \,.
\end{align}
Namely, we verify at the two-loop order and symbol level,
\begin{align}
 \log \left( \frac{F_6}{g^2 F_6^{(0)}}\right)  \sim g^2\, \frac{F^{(1)}_6}{F^{(0)}_6} + O(g^6) \,.\label{eq:logMR}
\end{align}
Previously, a similar behavior of the four-cusp and five-cusp Wilson loops with a Lagrangian insertion  was observed in the Regge/multi-Regge limit~\cite{Chicherin:2022zxo}. In those cases, there are beyond-the-symbol terms in the logarithm \eqref{eq:logMR} that contribute at order $O(g^4)$.

\bibliographystyle{JHEP}
\bibliography{refs}

\providecommand{\href}[2]{#2}\begingroup\raggedright\begin{thebibliography}{100}

\bibitem{Wilson}
K.~G. Wilson, \emph{{Confinement of Quarks}}, \href{http://dx.doi.org/10.1103/PhysRevD.10.2445}{\emph{Phys. Rev. D} {\bf 10} (1974) 2445--2459}.

\bibitem{Arkani-Hamed:2008owk}
N.~Arkani-Hamed, F.~Cachazo and J.~Kaplan, \emph{{What is the Simplest Quantum Field Theory?}}, \href{http://dx.doi.org/10.1007/JHEP09(2010)016}{\emph{JHEP} {\bf 09} (2010) 016}, [\href{http://arxiv.org/abs/0808.1446}{{\tt 0808.1446}}].

\bibitem{Arkani-Hamed:2022rwr}
N.~Arkani-Hamed, L.~J. Dixon, A.~J. McLeod, M.~Spradlin, J.~Trnka and A.~Volovich, \emph{{Solving Scattering in $N$ = 4 Super-Yang-Mills Theory}},  \href{http://arxiv.org/abs/2207.10636}{{\tt 2207.10636}}.

\bibitem{Travaglini:2022uwo}
G.~Travaglini et~al., \emph{{The SAGEX review on scattering amplitudes}}, \href{http://dx.doi.org/10.1088/1751-8121/ac8380}{\emph{J. Phys. A} {\bf 55} (2022) 443001}, [\href{http://arxiv.org/abs/2203.13011}{{\tt 2203.13011}}].

\bibitem{Alday:2007hr}
L.~F. Alday and J.~M. Maldacena, \emph{{Gluon scattering amplitudes at strong coupling}}, \href{http://dx.doi.org/10.1088/1126-6708/2007/06/064}{\emph{JHEP} {\bf 06} (2007) 064}, [\href{http://arxiv.org/abs/0705.0303}{{\tt 0705.0303}}].

\bibitem{Drummond:2007aua}
J.~M. Drummond, G.~P. Korchemsky and E.~Sokatchev, \emph{{Conformal properties of four-gluon planar amplitudes and Wilson loops}}, \href{http://dx.doi.org/10.1016/j.nuclphysb.2007.11.041}{\emph{Nucl. Phys. B} {\bf 795} (2008) 385--408}, [\href{http://arxiv.org/abs/0707.0243}{{\tt 0707.0243}}].

\bibitem{Brandhuber:2007yx}
A.~Brandhuber, P.~Heslop and G.~Travaglini, \emph{{MHV amplitudes in N=4 super Yang-Mills and Wilson loops}}, \href{http://dx.doi.org/10.1016/j.nuclphysb.2007.11.002}{\emph{Nucl. Phys. B} {\bf 794} (2008) 231--243}, [\href{http://arxiv.org/abs/0707.1153}{{\tt 0707.1153}}].

\bibitem{Mason:2010yk}
L.~J. Mason and D.~Skinner, \emph{{The Complete Planar S-matrix of N=4 SYM as a Wilson Loop in Twistor Space}}, \href{http://dx.doi.org/10.1007/JHEP12(2010)018}{\emph{JHEP} {\bf 12} (2010) 018}, [\href{http://arxiv.org/abs/1009.2225}{{\tt 1009.2225}}].

\bibitem{Caron-Huot:2010ryg}
S.~Caron-Huot, \emph{{Notes on the scattering amplitude / Wilson loop duality}}, \href{http://dx.doi.org/10.1007/JHEP07(2011)058}{\emph{JHEP} {\bf 07} (2011) 058}, [\href{http://arxiv.org/abs/1010.1167}{{\tt 1010.1167}}].

\bibitem{Drummond:2008vq}
J.~M. Drummond, J.~Henn, G.~P. Korchemsky and E.~Sokatchev, \emph{{Dual superconformal symmetry of scattering amplitudes in N=4 super-Yang-Mills theory}}, \href{http://dx.doi.org/10.1016/j.nuclphysb.2009.11.022}{\emph{Nucl. Phys. B} {\bf 828} (2010) 317--374}, [\href{http://arxiv.org/abs/0807.1095}{{\tt 0807.1095}}].

\bibitem{Drummond:2009fd}
J.~M. Drummond, J.~M. Henn and J.~Plefka, \emph{{Yangian symmetry of scattering amplitudes in N=4 super Yang-Mills theory}}, \href{http://dx.doi.org/10.1088/1126-6708/2009/05/046}{\emph{JHEP} {\bf 05} (2009) 046}, [\href{http://arxiv.org/abs/0902.2987}{{\tt 0902.2987}}].

\bibitem{Arkani-Hamed:2012zlh}
N.~Arkani-Hamed, J.~L. Bourjaily, F.~Cachazo, A.~B. Goncharov, A.~Postnikov and J.~Trnka, \emph{{Grassmannian Geometry of Scattering Amplitudes}}.
\newblock Cambridge University Press, 4, 2016.
\newblock 10.1017/CBO9781316091548.

\bibitem{Arkani-Hamed:2010zjl}
N.~Arkani-Hamed, J.~L. Bourjaily, F.~Cachazo, S.~Caron-Huot and J.~Trnka, \emph{{The All-Loop Integrand For Scattering Amplitudes in Planar N=4 SYM}}, \href{http://dx.doi.org/10.1007/JHEP01(2011)041}{\emph{JHEP} {\bf 01} (2011) 041}, [\href{http://arxiv.org/abs/1008.2958}{{\tt 1008.2958}}].

\bibitem{Arkani-Hamed:2013jha}
N.~Arkani-Hamed and J.~Trnka, \emph{{The Amplituhedron}}, \href{http://dx.doi.org/10.1007/JHEP10(2014)030}{\emph{JHEP} {\bf 10} (2014) 030}, [\href{http://arxiv.org/abs/1312.2007}{{\tt 1312.2007}}].

\bibitem{Arkani-Hamed:2013kca}
N.~Arkani-Hamed and J.~Trnka, \emph{{Into the Amplituhedron}}, \href{http://dx.doi.org/10.1007/JHEP12(2014)182}{\emph{JHEP} {\bf 12} (2014) 182}, [\href{http://arxiv.org/abs/1312.7878}{{\tt 1312.7878}}].

\bibitem{Drummond:2007au}
J.~M. Drummond, J.~Henn, G.~P. Korchemsky and E.~Sokatchev, \emph{{Conformal Ward identities for Wilson loops and a test of the duality with gluon amplitudes}}, \href{http://dx.doi.org/10.1016/j.nuclphysb.2009.10.013}{\emph{Nucl. Phys. B} {\bf 826} (2010) 337--364}, [\href{http://arxiv.org/abs/0712.1223}{{\tt 0712.1223}}].

\bibitem{Bern:2005iz}
Z.~Bern, L.~J. Dixon and V.~A. Smirnov, \emph{{Iteration of planar amplitudes in maximally supersymmetric Yang-Mills theory at three loops and beyond}}, \href{http://dx.doi.org/10.1103/PhysRevD.72.085001}{\emph{Phys. Rev. D} {\bf 72} (2005) 085001}, [\href{http://arxiv.org/abs/hep-th/0505205}{{\tt hep-th/0505205}}].

\bibitem{Goncharov:2010jf}
A.~B. Goncharov, M.~Spradlin, C.~Vergu and A.~Volovich, \emph{{Classical Polylogarithms for Amplitudes and Wilson Loops}}, \href{http://dx.doi.org/10.1103/PhysRevLett.105.151605}{\emph{Phys. Rev. Lett.} {\bf 105} (2010) 151605}, [\href{http://arxiv.org/abs/1006.5703}{{\tt 1006.5703}}].

\bibitem{Duhr:2011zq}
C.~Duhr, H.~Gangl and J.~R. Rhodes, \emph{{From polygons and symbols to polylogarithmic functions}}, \href{http://dx.doi.org/10.1007/JHEP10(2012)075}{\emph{JHEP} {\bf 10} (2012) 075}, [\href{http://arxiv.org/abs/1110.0458}{{\tt 1110.0458}}].

\bibitem{Golden:2013xva}
J.~Golden, A.~B. Goncharov, M.~Spradlin, C.~Vergu and A.~Volovich, \emph{{Motivic Amplitudes and Cluster Coordinates}}, \href{http://dx.doi.org/10.1007/JHEP01(2014)091}{\emph{JHEP} {\bf 01} (2014) 091}, [\href{http://arxiv.org/abs/1305.1617}{{\tt 1305.1617}}].

\bibitem{Dixon:2011pw}
L.~J. Dixon, J.~M. Drummond and J.~M. Henn, \emph{{Bootstrapping the three-loop hexagon}}, \href{http://dx.doi.org/10.1007/JHEP11(2011)023}{\emph{JHEP} {\bf 11} (2011) 023}, [\href{http://arxiv.org/abs/1108.4461}{{\tt 1108.4461}}].

\bibitem{Caron-Huot:2016owq}
S.~Caron-Huot, L.~J. Dixon, A.~McLeod and M.~von Hippel, \emph{{Bootstrapping a Five-Loop Amplitude Using Steinmann Relations}}, \href{http://dx.doi.org/10.1103/PhysRevLett.117.241601}{\emph{Phys. Rev. Lett.} {\bf 117} (2016) 241601}, [\href{http://arxiv.org/abs/1609.00669}{{\tt 1609.00669}}].

\bibitem{Dixon:2016nkn}
L.~J. Dixon, J.~Drummond, T.~Harrington, A.~J. McLeod, G.~Papathanasiou and M.~Spradlin, \emph{{Heptagons from the Steinmann Cluster Bootstrap}}, \href{http://dx.doi.org/10.1007/JHEP02(2017)137}{\emph{JHEP} {\bf 02} (2017) 137}, [\href{http://arxiv.org/abs/1612.08976}{{\tt 1612.08976}}].

\bibitem{Drummond:2018caf}
J.~Drummond, J.~Foster, {\"{O}}.~G{\"{u}}rdo{\u{g}}an and G.~Papathanasiou, \emph{{Cluster adjacency and the four-loop NMHV heptagon}}, \href{http://dx.doi.org/10.1007/JHEP03(2019)087}{\emph{JHEP} {\bf 03} (2019) 087}, [\href{http://arxiv.org/abs/1812.04640}{{\tt 1812.04640}}].

\bibitem{Caron-Huot:2019vjl}
S.~Caron-Huot, L.~J. Dixon, F.~Dulat, M.~von Hippel, A.~J. McLeod and G.~Papathanasiou, \emph{{Six-Gluon amplitudes in planar $ \mathcal{N} $ = 4 super-Yang-Mills theory at six and seven loops}}, \href{http://dx.doi.org/10.1007/JHEP08(2019)016}{\emph{JHEP} {\bf 08} (2019) 016}, [\href{http://arxiv.org/abs/1903.10890}{{\tt 1903.10890}}].

\bibitem{Caron-Huot:2019bsq}
S.~Caron-Huot, L.~J. Dixon, F.~Dulat, M.~von Hippel, A.~J. McLeod and G.~Papathanasiou, \emph{{The Cosmic Galois Group and Extended Steinmann Relations for Planar $\mathcal{N} = 4$ SYM Amplitudes}}, \href{http://dx.doi.org/10.1007/JHEP09(2019)061}{\emph{JHEP} {\bf 09} (2019) 061}, [\href{http://arxiv.org/abs/1906.07116}{{\tt 1906.07116}}].

\bibitem{Caron-Huot:2020bkp}
S.~Caron-Huot, L.~J. Dixon, J.~M. Drummond, F.~Dulat, J.~Foster, O.~G\"urdo\u{g}an, M.~von Hippel, A.~J. McLeod and G.~Papathanasiou, \emph{{The Steinmann Cluster Bootstrap for $N$ = 4 Super Yang-Mills Amplitudes}}, \href{http://dx.doi.org/10.22323/1.376.0003}{\emph{PoS} {\bf CORFU2019} (2020) 003}, [\href{http://arxiv.org/abs/2005.06735}{{\tt 2005.06735}}].

\bibitem{Caron-Huot:2011zgw}
S.~Caron-Huot, \emph{{Superconformal symmetry and two-loop amplitudes in planar N=4 super Yang-Mills}}, \href{http://dx.doi.org/10.1007/JHEP12(2011)066}{\emph{JHEP} {\bf 12} (2011) 066}, [\href{http://arxiv.org/abs/1105.5606}{{\tt 1105.5606}}].

\bibitem{Caron-Huot:2011dec}
S.~Caron-Huot and S.~He, \emph{{Jumpstarting the All-Loop S-Matrix of Planar N=4 Super Yang-Mills}}, \href{http://dx.doi.org/10.1007/JHEP07(2012)174}{\emph{JHEP} {\bf 07} (2012) 174}, [\href{http://arxiv.org/abs/1112.1060}{{\tt 1112.1060}}].

\bibitem{Alday:2010ku}
L.~F. Alday, D.~Gaiotto, J.~Maldacena, A.~Sever and P.~Vieira, \emph{{An Operator Product Expansion for Polygonal null Wilson Loops}}, \href{http://dx.doi.org/10.1007/JHEP04(2011)088}{\emph{JHEP} {\bf 04} (2011) 088}, [\href{http://arxiv.org/abs/1006.2788}{{\tt 1006.2788}}].

\bibitem{Basso:2013vsa}
B.~Basso, A.~Sever and P.~Vieira, \emph{{Spacetime and Flux Tube S-Matrices at Finite Coupling for N=4 Supersymmetric Yang-Mills Theory}}, \href{http://dx.doi.org/10.1103/PhysRevLett.111.091602}{\emph{Phys. Rev. Lett.} {\bf 111} (2013) 091602}, [\href{http://arxiv.org/abs/1303.1396}{{\tt 1303.1396}}].

\bibitem{Basso:2013aha}
B.~Basso, A.~Sever and P.~Vieira, \emph{{Space-time S-matrix and Flux tube S-matrix II. Extracting and Matching Data}}, \href{http://dx.doi.org/10.1007/JHEP01(2014)008}{\emph{JHEP} {\bf 01} (2014) 008}, [\href{http://arxiv.org/abs/1306.2058}{{\tt 1306.2058}}].

\bibitem{Broadhurst:1993ib}
D.~J. Broadhurst, \emph{{Summation of an infinite series of ladder diagrams}}, \href{http://dx.doi.org/10.1016/0370-2693(93)90202-S}{\emph{Phys. Lett. B} {\bf 307} (1993) 132--139}.

\bibitem{Alday:2011ga}
L.~F. Alday, E.~I. Buchbinder and A.~A. Tseytlin, \emph{{Correlation function of null polygonal Wilson loops with local operators}}, \href{http://dx.doi.org/10.1007/JHEP09(2011)034}{\emph{JHEP} {\bf 09} (2011) 034}, [\href{http://arxiv.org/abs/1107.5702}{{\tt 1107.5702}}].

\bibitem{Alday:2012hy}
L.~F. Alday, P.~Heslop and J.~Sikorowski, \emph{{Perturbative correlation functions of null Wilson loops and local operators}}, \href{http://dx.doi.org/10.1007/JHEP03(2013)074}{\emph{JHEP} {\bf 03} (2013) 074}, [\href{http://arxiv.org/abs/1207.4316}{{\tt 1207.4316}}].

\bibitem{Alday:2013ip}
L.~F. Alday, J.~M. Henn and J.~Sikorowski, \emph{{Higher loop mixed correlators in N=4 SYM}}, \href{http://dx.doi.org/10.1007/JHEP03(2013)058}{\emph{JHEP} {\bf 03} (2013) 058}, [\href{http://arxiv.org/abs/1301.0149}{{\tt 1301.0149}}].

\bibitem{Engelund:2011fg}
O.~T. Engelund and R.~Roiban, \emph{{On correlation functions of Wilson loops, local and non-local operators}}, \href{http://dx.doi.org/10.1007/JHEP05(2012)158}{\emph{JHEP} {\bf 05} (2012) 158}, [\href{http://arxiv.org/abs/1110.0758}{{\tt 1110.0758}}].

\bibitem{Engelund:2012re}
O.~T. Engelund and R.~Roiban, \emph{{Correlation functions of local composite operators from generalized unitarity}}, \href{http://dx.doi.org/10.1007/JHEP03(2013)172}{\emph{JHEP} {\bf 03} (2013) 172}, [\href{http://arxiv.org/abs/1209.0227}{{\tt 1209.0227}}].

\bibitem{Henn:2019swt}
J.~M. Henn, G.~P. Korchemsky and B.~Mistlberger, \emph{{The full four-loop cusp anomalous dimension in $\mathcal{N}=4$ super Yang-Mills and QCD}}, \href{http://dx.doi.org/10.1007/JHEP04(2020)018}{\emph{JHEP} {\bf 04} (2020) 018}, [\href{http://arxiv.org/abs/1911.10174}{{\tt 1911.10174}}].

\bibitem{Chicherin:2022bov}
D.~Chicherin and J.~M. Henn, \emph{{Symmetry properties of Wilson loops with a Lagrangian insertion}}, \href{http://dx.doi.org/10.1007/JHEP07(2022)057}{\emph{JHEP} {\bf 07} (2022) 057}, [\href{http://arxiv.org/abs/2202.05596}{{\tt 2202.05596}}].

\bibitem{Chicherin:2022zxo}
D.~Chicherin and J.~Henn, \emph{{Pentagon Wilson loop with Lagrangian insertion at two loops in $ \mathcal{N} $ = 4 super Yang-Mills theory}}, \href{http://dx.doi.org/10.1007/JHEP07(2022)038}{\emph{JHEP} {\bf 07} (2022) 038}, [\href{http://arxiv.org/abs/2204.00329}{{\tt 2204.00329}}].

\bibitem{Bargheer:2024hfx}
T.~Bargheer, C.~Bercini, B.~Fernandes, V.~Gon\c{c}alves and J.~Mann, \emph{{Wilson Loops with Lagrangians: Large-Spin Operator Product Expansion and Cusp Anomalous Dimension Dictionary}}, \href{http://dx.doi.org/10.1103/PhysRevLett.134.141601}{\emph{Phys. Rev. Lett.} {\bf 134} (2025) 141601}, [\href{http://arxiv.org/abs/2406.04294}{{\tt 2406.04294}}].

\bibitem{Chicherin:2024hes}
D.~Chicherin, J.~Henn, J.~Trnka and S.-Q. Zhang, \emph{{Positivity properties of five-point two-loop Wilson loops with Lagrangian insertion}},  \href{http://arxiv.org/abs/2410.11456}{{\tt 2410.11456}}.

\bibitem{Eden:2011yp}
B.~Eden, P.~Heslop, G.~P. Korchemsky and E.~Sokatchev, \emph{{The super-correlator/super-amplitude duality: Part I}}, \href{http://dx.doi.org/10.1016/j.nuclphysb.2012.12.015}{\emph{Nucl. Phys. B} {\bf 869} (2013) 329--377}, [\href{http://arxiv.org/abs/1103.3714}{{\tt 1103.3714}}].

\bibitem{Cachazo:2008vp}
F.~Cachazo, \emph{{Sharpening The Leading Singularity}},  \href{http://arxiv.org/abs/0803.1988}{{\tt 0803.1988}}.

\bibitem{Bullimore:2009cb}
M.~Bullimore, L.~J. Mason and D.~Skinner, \emph{{Twistor-Strings, Grassmannians and Leading Singularities}}, \href{http://dx.doi.org/10.1007/JHEP03(2010)070}{\emph{JHEP} {\bf 03} (2010) 070}, [\href{http://arxiv.org/abs/0912.0539}{{\tt 0912.0539}}].

\bibitem{Arkani-Hamed:2010pyv}
N.~Arkani-Hamed, J.~L. Bourjaily, F.~Cachazo and J.~Trnka, \emph{{Local Integrals for Planar Scattering Amplitudes}}, \href{http://dx.doi.org/10.1007/JHEP06(2012)125}{\emph{JHEP} {\bf 06} (2012) 125}, [\href{http://arxiv.org/abs/1012.6032}{{\tt 1012.6032}}].

\bibitem{Dixon:2014xca}
L.~J. Dixon, J.~M. Drummond, C.~Duhr, M.~von Hippel and J.~Pennington, \emph{{Bootstrapping six-gluon scattering in planar N=4 super-Yang-Mills theory}}, \href{http://dx.doi.org/10.22323/1.211.0077}{\emph{PoS} {\bf LL2014} (2014) 077}, [\href{http://arxiv.org/abs/1407.4724}{{\tt 1407.4724}}].

\bibitem{Dixon:2014iba}
L.~J. Dixon and M.~von Hippel, \emph{{Bootstrapping an NMHV amplitude through three loops}}, \href{http://dx.doi.org/10.1007/JHEP10(2014)065}{\emph{JHEP} {\bf 10} (2014) 065}, [\href{http://arxiv.org/abs/1408.1505}{{\tt 1408.1505}}].

\bibitem{Drummond:2014ffa}
J.~M. Drummond, G.~Papathanasiou and M.~Spradlin, \emph{{A Symbol of Uniqueness: The Cluster Bootstrap for the 3-Loop MHV Heptagon}}, \href{http://dx.doi.org/10.1007/JHEP03(2015)072}{\emph{JHEP} {\bf 03} (2015) 072}, [\href{http://arxiv.org/abs/1412.3763}{{\tt 1412.3763}}].

\bibitem{Dixon:2015iva}
L.~J. Dixon, M.~von Hippel and A.~J. McLeod, \emph{{The four-loop six-gluon NMHV ratio function}}, \href{http://dx.doi.org/10.1007/JHEP01(2016)053}{\emph{JHEP} {\bf 01} (2016) 053}, [\href{http://arxiv.org/abs/1509.08127}{{\tt 1509.08127}}].

\bibitem{Dixon:2020bbt}
L.~J. Dixon, A.~J. McLeod and M.~Wilhelm, \emph{{A Three-Point Form Factor Through Five Loops}}, \href{http://dx.doi.org/10.1007/JHEP04(2021)147}{\emph{JHEP} {\bf 04} (2021) 147}, [\href{http://arxiv.org/abs/2012.12286}{{\tt 2012.12286}}].

\bibitem{Dixon:2022rse}
L.~J. Dixon, O.~Gurdogan, A.~J. McLeod and M.~Wilhelm, \emph{{Bootstrapping a stress-tensor form factor through eight loops}}, \href{http://dx.doi.org/10.1007/JHEP07(2022)153}{\emph{JHEP} {\bf 07} (2022) 153}, [\href{http://arxiv.org/abs/2204.11901}{{\tt 2204.11901}}].

\bibitem{Dixon:2022xqh}
L.~J. Dixon, O.~G\"urdo\u{g}an, Y.-T. Liu, A.~J. McLeod and M.~Wilhelm, \emph{{Antipodal Self-Duality for a Four-Particle Form Factor}}, \href{http://dx.doi.org/10.1103/PhysRevLett.130.111601}{\emph{Phys. Rev. Lett.} {\bf 130} (2023) 111601}, [\href{http://arxiv.org/abs/2212.02410}{{\tt 2212.02410}}].

\bibitem{Dixon:2023kop}
L.~J. Dixon and Y.-T. Liu, \emph{{An eight loop amplitude via antipodal duality}}, \href{http://dx.doi.org/10.1007/JHEP09(2023)098}{\emph{JHEP} {\bf 09} (2023) 098}, [\href{http://arxiv.org/abs/2308.08199}{{\tt 2308.08199}}].

\bibitem{Basso:2024hlx}
B.~Basso, L.~J. Dixon and A.~G. Tumanov, \emph{{The three-point form factor of Tr \ensuremath{\phi}$^{3}$ to six loops}}, \href{http://dx.doi.org/10.1007/JHEP02(2025)034}{\emph{JHEP} {\bf 02} (2025) 034}, [\href{http://arxiv.org/abs/2410.22402}{{\tt 2410.22402}}].

\bibitem{Caola:2020dfu}
F.~Caola, A.~Von~Manteuffel and L.~Tancredi, \emph{{Diphoton Amplitudes in Three-Loop Quantum Chromodynamics}}, \href{http://dx.doi.org/10.1103/PhysRevLett.126.112004}{\emph{Phys. Rev. Lett.} {\bf 126} (2021) 112004}, [\href{http://arxiv.org/abs/2011.13946}{{\tt 2011.13946}}].

\bibitem{Caola:2021izf}
F.~Caola, A.~Chakraborty, G.~Gambuti, A.~von Manteuffel and L.~Tancredi, \emph{{Three-Loop Gluon Scattering in QCD and the Gluon Regge Trajectory}}, \href{http://dx.doi.org/10.1103/PhysRevLett.128.212001}{\emph{Phys. Rev. Lett.} {\bf 128} (2022) 212001}, [\href{http://arxiv.org/abs/2112.11097}{{\tt 2112.11097}}].

\bibitem{Caola:2021rqz}
F.~Caola, A.~Chakraborty, G.~Gambuti, A.~von Manteuffel and L.~Tancredi, \emph{{Three-loop helicity amplitudes for four-quark scattering in massless QCD}}, \href{http://dx.doi.org/10.1007/JHEP10(2021)206}{\emph{JHEP} {\bf 10} (2021) 206}, [\href{http://arxiv.org/abs/2108.00055}{{\tt 2108.00055}}].

\bibitem{Abreu:2018zmy}
S.~Abreu, J.~Dormans, F.~Febres~Cordero, H.~Ita and B.~Page, \emph{{Analytic Form of Planar Two-Loop Five-Gluon Scattering Amplitudes in QCD}}, \href{http://dx.doi.org/10.1103/PhysRevLett.122.082002}{\emph{Phys. Rev. Lett.} {\bf 122} (2019) 082002}, [\href{http://arxiv.org/abs/1812.04586}{{\tt 1812.04586}}].

\bibitem{Abreu:2019odu}
S.~Abreu, J.~Dormans, F.~Febres~Cordero, H.~Ita, B.~Page and V.~Sotnikov, \emph{{Analytic Form of the Planar Two-Loop Five-Parton Scattering Amplitudes in QCD}}, \href{http://dx.doi.org/10.1007/JHEP05(2019)084}{\emph{JHEP} {\bf 05} (2019) 084}, [\href{http://arxiv.org/abs/1904.00945}{{\tt 1904.00945}}].

\bibitem{Abreu:2021oya}
S.~Abreu, F.~Febres~Cordero, H.~Ita, B.~Page and V.~Sotnikov, \emph{{Leading-color two-loop QCD corrections for three-jet production at hadron colliders}}, \href{http://dx.doi.org/10.1007/JHEP07(2021)095}{\emph{JHEP} {\bf 07} (2021) 095}, [\href{http://arxiv.org/abs/2102.13609}{{\tt 2102.13609}}].

\bibitem{Badger:2013gxa}
S.~Badger, H.~Frellesvig and Y.~Zhang, \emph{{A Two-Loop Five-Gluon Helicity Amplitude in QCD}}, \href{http://dx.doi.org/10.1007/JHEP12(2013)045}{\emph{JHEP} {\bf 12} (2013) 045}, [\href{http://arxiv.org/abs/1310.1051}{{\tt 1310.1051}}].

\bibitem{Badger:2015lda}
S.~Badger, G.~Mogull, A.~Ochirov and D.~O'Connell, \emph{{A Complete Two-Loop, Five-Gluon Helicity Amplitude in Yang-Mills Theory}}, \href{http://dx.doi.org/10.1007/JHEP10(2015)064}{\emph{JHEP} {\bf 10} (2015) 064}, [\href{http://arxiv.org/abs/1507.08797}{{\tt 1507.08797}}].

\bibitem{Badger:2018enw}
S.~Badger, C.~Br\o{}nnum-Hansen, H.~B. Hartanto and T.~Peraro, \emph{{Analytic helicity amplitudes for two-loop five-gluon scattering: the single-minus case}}, \href{http://dx.doi.org/10.1007/JHEP01(2019)186}{\emph{JHEP} {\bf 01} (2019) 186}, [\href{http://arxiv.org/abs/1811.11699}{{\tt 1811.11699}}].

\bibitem{Badger:2019djh}
S.~Badger, D.~Chicherin, T.~Gehrmann, G.~Heinrich, J.~M. Henn, T.~Peraro, P.~Wasser, Y.~Zhang and S.~Zoia, \emph{{Analytic form of the full two-loop five-gluon all-plus helicity amplitude}}, \href{http://dx.doi.org/10.1103/PhysRevLett.123.071601}{\emph{Phys. Rev. Lett.} {\bf 123} (2019) 071601}, [\href{http://arxiv.org/abs/1905.03733}{{\tt 1905.03733}}].

\bibitem{Badger:2023eqz}
S.~Badger, J.~Henn, J.~C. Plefka and S.~Zoia, \emph{{Scattering Amplitudes in Quantum Field Theory}}, \href{http://dx.doi.org/10.1007/978-3-031-46987-9}{\emph{Lect. Notes Phys.} {\bf 1021} (2024) pp.}, [\href{http://arxiv.org/abs/2306.05976}{{\tt 2306.05976}}].

\bibitem{Agarwal:2023suw}
B.~Agarwal, F.~Buccioni, F.~Devoto, G.~Gambuti, A.~von Manteuffel and L.~Tancredi, \emph{{Five-parton scattering in QCD at two loops}}, \href{http://dx.doi.org/10.1103/PhysRevD.109.094025}{\emph{Phys. Rev. D} {\bf 109} (2024) 094025}, [\href{http://arxiv.org/abs/2311.09870}{{\tt 2311.09870}}].

\bibitem{DeLaurentis:2023nss}
G.~De~Laurentis, H.~Ita, M.~Klinkert and V.~Sotnikov, \emph{{Double-virtual NNLO QCD corrections for five-parton scattering. I. The gluon channel}}, \href{http://dx.doi.org/10.1103/PhysRevD.109.094023}{\emph{Phys. Rev. D} {\bf 109} (2024) 094023}, [\href{http://arxiv.org/abs/2311.10086}{{\tt 2311.10086}}].

\bibitem{DeLaurentis:2023izi}
G.~De~Laurentis, H.~Ita and V.~Sotnikov, \emph{{Double-virtual NNLO QCD corrections for five-parton scattering. II. The quark channels}}, \href{http://dx.doi.org/10.1103/PhysRevD.109.094024}{\emph{Phys. Rev. D} {\bf 109} (2024) 094024}, [\href{http://arxiv.org/abs/2311.18752}{{\tt 2311.18752}}].

\bibitem{Henn:2016jdu}
J.~M. Henn and B.~Mistlberger, \emph{{Four-Gluon Scattering at Three Loops, Infrared Structure, and the Regge Limit}}, \href{http://dx.doi.org/10.1103/PhysRevLett.117.171601}{\emph{Phys. Rev. Lett.} {\bf 117} (2016) 171601}, [\href{http://arxiv.org/abs/1608.00850}{{\tt 1608.00850}}].

\bibitem{Abreu:2018aqd}
S.~Abreu, L.~J. Dixon, E.~Herrmann, B.~Page and M.~Zeng, \emph{{The two-loop five-point amplitude in $\mathcal{N} =4$ super-Yang-Mills theory}}, \href{http://dx.doi.org/10.1103/PhysRevLett.122.121603}{\emph{Phys. Rev. Lett.} {\bf 122} (2019) 121603}, [\href{http://arxiv.org/abs/1812.08941}{{\tt 1812.08941}}].

\bibitem{Chicherin:2018yne}
D.~Chicherin, T.~Gehrmann, J.~M. Henn, P.~Wasser, Y.~Zhang and S.~Zoia, \emph{{Analytic result for a two-loop five-particle amplitude}}, \href{http://dx.doi.org/10.1103/PhysRevLett.122.121602}{\emph{Phys. Rev. Lett.} {\bf 122} (2019) 121602}, [\href{http://arxiv.org/abs/1812.11057}{{\tt 1812.11057}}].

\bibitem{Henn:2025xrc}
J.~Henn, A.~Matija\v{s}i\'c, J.~Miczajka, T.~Peraro, Y.~Xu and Y.~Zhang, \emph{{Complete function space for planar two-loop six-particle scattering amplitudes}},  \href{http://arxiv.org/abs/2501.01847}{{\tt 2501.01847}}.

\bibitem{Abreu:2024fei}
S.~Abreu, P.~F. Monni, B.~Page and J.~Usovitsch, \emph{{Planar Six-Point Feynman Integrals for Four-Dimensional Gauge Theories}},  \href{http://arxiv.org/abs/2412.19884}{{\tt 2412.19884}}.

\bibitem{Henn:2013pwa}
J.~M. Henn, \emph{{Multiloop integrals in dimensional regularization made simple}}, \href{http://dx.doi.org/10.1103/PhysRevLett.110.251601}{\emph{Phys. Rev. Lett.} {\bf 110} (2013) 251601}, [\href{http://arxiv.org/abs/1304.1806}{{\tt 1304.1806}}].

\bibitem{Brown:2025plq}
T.~V. Brown, J.~M. Henn, E.~Mazzucchelli and J.~Trnka, \emph{{All-loop Leading Singularities of Wilson Loops}},  \href{http://arxiv.org/abs/2503.17185}{{\tt 2503.17185}}.

\bibitem{Guo:2021bym}
Y.~Guo, L.~Wang and G.~Yang, \emph{{Bootstrapping a Two-Loop Four-Point Form Factor}}, \href{http://dx.doi.org/10.1103/PhysRevLett.127.151602}{\emph{Phys. Rev. Lett.} {\bf 127} (2021) 151602}, [\href{http://arxiv.org/abs/2106.01374}{{\tt 2106.01374}}].

\bibitem{Hodges:2009hk}
A.~Hodges, \emph{{Eliminating spurious poles from gauge-theoretic amplitudes}}, \href{http://dx.doi.org/10.1007/JHEP05(2013)135}{\emph{JHEP} {\bf 05} (2013) 135}, [\href{http://arxiv.org/abs/0905.1473}{{\tt 0905.1473}}].

\bibitem{goncharov2005galois}
A.~B. Goncharov et~al., \emph{Galois symmetries of fundamental groupoids and noncommutative geometry}, {\emph{Duke Mathematical Journal} {\bf 128} (2005) 209--284}.

\bibitem{Duhr:2019tlz}
C.~Duhr and F.~Dulat, \emph{{PolyLogTools \textemdash{} polylogs for the masses}}, \href{http://dx.doi.org/10.1007/JHEP08(2019)135}{\emph{JHEP} {\bf 08} (2019) 135}, [\href{http://arxiv.org/abs/1904.07279}{{\tt 1904.07279}}].

\bibitem{Dixon:2021tdw}
L.~J. Dixon, O.~Gurdogan, A.~J. McLeod and M.~Wilhelm, \emph{{Folding Amplitudes into Form Factors: An Antipodal Duality}}, \href{http://dx.doi.org/10.1103/PhysRevLett.128.111602}{\emph{Phys. Rev. Lett.} {\bf 128} (2022) 111602}, [\href{http://arxiv.org/abs/2112.06243}{{\tt 2112.06243}}].

\bibitem{Henn:2021cyv}
J.~Henn, T.~Peraro, Y.~Xu and Y.~Zhang, \emph{{A first look at the function space for planar two-loop six-particle Feynman integrals}}, \href{http://dx.doi.org/10.1007/JHEP03(2022)056}{\emph{JHEP} {\bf 03} (2022) 056}, [\href{http://arxiv.org/abs/2112.10605}{{\tt 2112.10605}}].

\bibitem{Henn:2022ydo}
J.~M. Henn, A.~Matija\v{s}i\'c and J.~Miczajka, \emph{{One-loop hexagon integral to higher orders in the dimensional regulator}}, \href{http://dx.doi.org/10.1007/JHEP01(2023)096}{\emph{JHEP} {\bf 01} (2023) 096}, [\href{http://arxiv.org/abs/2210.13505}{{\tt 2210.13505}}].

\bibitem{Henn:2024ngj}
J.~M. Henn, A.~Matija\v{s}i\'c, J.~Miczajka, T.~Peraro, Y.~Xu and Y.~Zhang, \emph{{A computation of two-loop six-point Feynman integrals in dimensional regularization}}, \href{http://dx.doi.org/10.1007/JHEP08(2024)027}{\emph{JHEP} {\bf 08} (2024) 027}, [\href{http://arxiv.org/abs/2403.19742}{{\tt 2403.19742}}].

\bibitem{Kotikov:2004er}
A.~V. Kotikov, L.~N. Lipatov, A.~I. Onishchenko and V.~N. Velizhanin, \emph{{Three loop universal anomalous dimension of the Wilson operators in $N=4$ SUSY Yang-Mills model}}, \href{http://dx.doi.org/10.1016/j.physletb.2004.05.078}{\emph{Phys. Lett. B} {\bf 595} (2004) 521--529}, [\href{http://arxiv.org/abs/hep-th/0404092}{{\tt hep-th/0404092}}].

\bibitem{Hannesdottir:2024hke}
H.~S. Hannesdottir, A.~J. McLeod, M.~D. Schwartz and C.~Vergu, \emph{{Applications of the Landau bootstrap}}, \href{http://dx.doi.org/10.1103/PhysRevD.111.085003}{\emph{Phys. Rev. D} {\bf 111} (2025) 085003}, [\href{http://arxiv.org/abs/2410.02424}{{\tt 2410.02424}}].

\bibitem{He:2023umf}
S.~He, X.~Jiang, J.~Liu and Q.~Yang, \emph{{On symbology and differential equations of Feynman integrals from Schubert analysis}}, \href{http://dx.doi.org/10.1007/JHEP12(2023)140}{\emph{JHEP} {\bf 12} (2023) 140}, [\href{http://arxiv.org/abs/2309.16441}{{\tt 2309.16441}}].

\bibitem{Caron-Huot:2020vlo}
S.~Caron-Huot, D.~Chicherin, J.~Henn, Y.~Zhang and S.~Zoia, \emph{{Multi-Regge Limit of the Two-Loop Five-Point Amplitudes in $\mathcal{N} = 4$ Super Yang-Mills and $\mathcal{N} = 8$ Supergravity}}, \href{http://dx.doi.org/10.1007/JHEP10(2020)188}{\emph{JHEP} {\bf 10} (2020) 188}, [\href{http://arxiv.org/abs/2003.03120}{{\tt 2003.03120}}].

\bibitem{Henn:2024qjq}
J.~Henn, R.~Ma, Y.~Xu, K.~Yan, Y.~Zhang and H.~X. Zhu, \emph{{Two-Loop Spacelike Splitting Amplitude for N=4 Super-Yang-Mills Theory}},  \href{http://arxiv.org/abs/2406.14604}{{\tt 2406.14604}}.

\bibitem{Henn:2024qwe}
J.~Henn and P.~Raman, \emph{{Positivity properties of scattering amplitudes}}, \href{http://dx.doi.org/10.1007/JHEP04(2025)150}{\emph{JHEP} {\bf 04} (2025) 150}, [\href{http://arxiv.org/abs/2407.05755}{{\tt 2407.05755}}].

\bibitem{Regge:1959mz}
T.~Regge, \emph{{Introduction to complex orbital momenta}}, \href{http://dx.doi.org/10.1007/BF02728177}{\emph{Nuovo Cim.} {\bf 14} (1959) 951}.

\bibitem{DelDuca:2022skz}
V.~Del~Duca and L.~J. Dixon, \emph{{The SAGEX review on scattering amplitudes Chapter 15: The multi-Regge limit}}, \href{http://dx.doi.org/10.1088/1751-8121/ac845c}{\emph{J. Phys. A} {\bf 55} (2022) 443016}, [\href{http://arxiv.org/abs/2203.13026}{{\tt 2203.13026}}].

\bibitem{Dorn:2011gf}
H.~Dorn and S.~Wuttke, \emph{{Wilson loop remainder function for null polygons in the limit of self-crossing}}, \href{http://dx.doi.org/10.1007/JHEP05(2011)114}{\emph{JHEP} {\bf 05} (2011) 114}, [\href{http://arxiv.org/abs/1104.2469}{{\tt 1104.2469}}].

\bibitem{Dorn:2011ec}
H.~Dorn and S.~Wuttke, \emph{{Hexagon Remainder Function in the Limit of Self-Crossing up to three Loops}}, \href{http://dx.doi.org/10.1007/JHEP04(2012)023}{\emph{JHEP} {\bf 04} (2012) 023}, [\href{http://arxiv.org/abs/1111.6815}{{\tt 1111.6815}}].

\bibitem{Dixon:2016epj}
L.~J. Dixon and I.~Esterlis, \emph{{All orders results for self-crossing Wilson loops mimicking double parton scattering}}, \href{http://dx.doi.org/10.1007/JHEP07(2016)116}{\emph{JHEP} {\bf 07} (2016) 116}, [\href{http://arxiv.org/abs/1602.02107}{{\tt 1602.02107}}].

\bibitem{Zhang:2019vnm}
S.~He, Z.~Li and C.~Zhang, \emph{{Two-loop Octagons, Algebraic Letters and $\bar{Q}$ Equations}}, \href{http://dx.doi.org/10.1103/PhysRevD.101.061701}{\emph{Phys. Rev. D} {\bf 101} (2020) 061701}, [\href{http://arxiv.org/abs/1911.01290}{{\tt 1911.01290}}].

\bibitem{Henke:2019hve}
N.~Henke and G.~Papathanasiou, \emph{{How tropical are seven- and eight-particle amplitudes?}}, \href{http://dx.doi.org/10.1007/JHEP08(2020)005}{\emph{JHEP} {\bf 08} (2020) 005}, [\href{http://arxiv.org/abs/1912.08254}{{\tt 1912.08254}}].

\bibitem{Drummond:2019qjk}
J.~Drummond, J.~Foster, O.~G\"urdogan and C.~Kalousios, \emph{{Tropical Grassmannians, cluster algebras and scattering amplitudes}}, \href{http://dx.doi.org/10.1007/JHEP04(2020)146}{\emph{JHEP} {\bf 04} (2020) 146}, [\href{http://arxiv.org/abs/1907.01053}{{\tt 1907.01053}}].

\bibitem{Arkani-Hamed:2019rds}
N.~Arkani-Hamed, T.~Lam and M.~Spradlin, \emph{{Non-perturbative geometries for planar $ \mathcal{N} $ = 4 SYM amplitudes}}, \href{http://dx.doi.org/10.1007/JHEP03(2021)065}{\emph{JHEP} {\bf 03} (2021) 065}, [\href{http://arxiv.org/abs/1912.08222}{{\tt 1912.08222}}].

\bibitem{Early:2023tjj}
N.~Early and J.-R. Li, \emph{{Tropical geometry, quantum affine algebras, and scattering amplitudes}}, \href{http://dx.doi.org/10.1088/1751-8121/ad909b}{\emph{J. Phys. A} {\bf 57} (2024) 495201}, [\href{http://arxiv.org/abs/2303.05618}{{\tt 2303.05618}}].

\bibitem{Caron-Huot:2018dsv}
S.~Caron-Huot, L.~J. Dixon, M.~von Hippel, A.~J. McLeod and G.~Papathanasiou, \emph{{The Double Pentaladder Integral to All Orders}}, \href{http://dx.doi.org/10.1007/JHEP07(2018)170}{\emph{JHEP} {\bf 07} (2018) 170}, [\href{http://arxiv.org/abs/1806.01361}{{\tt 1806.01361}}].

\bibitem{Henn:2018cdp}
J.~Henn, E.~Herrmann and J.~Parra-Martinez, \emph{{Bootstrapping two-loop Feynman integrals for planar $ \mathcal{N}=4 $ sYM}}, \href{http://dx.doi.org/10.1007/JHEP10(2018)059}{\emph{JHEP} {\bf 10} (2018) 059}, [\href{http://arxiv.org/abs/1806.06072}{{\tt 1806.06072}}].

\bibitem{Chicherin:2020umh}
D.~Chicherin, J.~M. Henn and G.~Papathanasiou, \emph{{Cluster algebras for Feynman integrals}}, \href{http://dx.doi.org/10.1103/PhysRevLett.126.091603}{\emph{Phys. Rev. Lett.} {\bf 126} (2021) 091603}, [\href{http://arxiv.org/abs/2012.12285}{{\tt 2012.12285}}].

\bibitem{He:2021esx}
S.~He, Z.~Li and Q.~Yang, \emph{{Notes on cluster algebras and some all-loop Feynman integrals}}, \href{http://dx.doi.org/10.1007/JHEP06(2021)119}{\emph{JHEP} {\bf 06} (2021) 119}, [\href{http://arxiv.org/abs/2103.02796}{{\tt 2103.02796}}].

\bibitem{He:2021non}
S.~He, Z.~Li and Q.~Yang, \emph{{Truncated cluster algebras and Feynman integrals with algebraic letters}}, \href{http://dx.doi.org/10.1007/JHEP12(2021)110}{\emph{JHEP} {\bf 12} (2021) 110}, [\href{http://arxiv.org/abs/2106.09314}{{\tt 2106.09314}}].

\bibitem{Aliaj:2024zgp}
R.~Aliaj and G.~Papathanasiou, \emph{{An exceptional cluster algebra for Higgs plus jet production}}, \href{http://dx.doi.org/10.1007/JHEP01(2025)197}{\emph{JHEP} {\bf 01} (2025) 197}, [\href{http://arxiv.org/abs/2408.14544}{{\tt 2408.14544}}].

\bibitem{Bossinger:2022eiy}
L.~Bossinger, J.~M. Drummond and R.~Glew, \emph{{Adjacency for scattering amplitudes from the Gr\"obner fan}}, \href{http://dx.doi.org/10.1007/JHEP11(2023)002}{\emph{JHEP} {\bf 11} (2023) 002}, [\href{http://arxiv.org/abs/2212.08931}{{\tt 2212.08931}}].

\bibitem{Bossinger:2024apm}
L.~Bossinger and J.-R. Li, \emph{{Cluster structures on spinor helicity and momentum twistor varieties}},  \href{http://arxiv.org/abs/2408.14956}{{\tt 2408.14956}}.

\bibitem{Chicherin:2017dob}
D.~Chicherin, J.~Henn and V.~Mitev, \emph{{Bootstrapping pentagon functions}}, \href{http://dx.doi.org/10.1007/JHEP05(2018)164}{\emph{JHEP} {\bf 05} (2018) 164}, [\href{http://arxiv.org/abs/1712.09610}{{\tt 1712.09610}}].

\bibitem{Gehrmann:2018yef}
T.~Gehrmann, J.~M. Henn and N.~A. Lo~Presti, \emph{{Pentagon functions for massless planar scattering amplitudes}}, \href{http://dx.doi.org/10.1007/JHEP10(2018)103}{\emph{JHEP} {\bf 10} (2018) 103}, [\href{http://arxiv.org/abs/1807.09812}{{\tt 1807.09812}}].

\end{thebibliography}\endgroup
\end{document}